# *Topological theory of non-Hermitian photonic systems*

*Mário G. Silveirinha*[*]

[(1)] *University of Lisbon–Instituto Superior Técnico and Instituto de Telecomunicações, Avenida Rovisco Pais, 1, 1049-001 Lisboa, Portugal,* [mario.silveirinha@co.it.pt](mario.silveirinha@co.it.pt)

**Abstract**

Here, we develop a gauge-independent Green function approach to characterize the Chern invariants of generic non-Hermitian systems. It is shown that analogous to the Hermitian case, the Chern number can be expressed as an integral of the system Green function over a line parallel to the imaginary-frequency axis. The approach introduces in a natural way the "band-gaps" of non-Hermitian systems as the strips of the complex-frequency plane wherein the system Green function is analytical. We apply the developed theory to nonreciprocal electromagnetic continua, showing that the topological properties of gyrotropic materials are strongly robust to the effect of material loss. Furthermore, it is proven that the spectrum of a topological material cavity terminated with opaque-type walls must be gapless. This result suggests that the bulk-edge correspondence remains valid for a class of non-Hermitian systems.

---

[*] To whom correspondence should be addressed: E-mail: [mario.silveirinha@co.it.pt](mario.silveirinha@co.it.pt)

# I. Introduction

Topological matter and topological systems have quite unique and often intriguing properties, which can lead to novel physical effects and phenomena [1-9]. The topological classification of materials was initially developed for Hermitian systems described by some self-adjoint operator. Recently, it was shown that non-Hermitian systems, for example systems with material absorption or material gain, can also have topological properties [10-24]. The topological Chern invariants of non-Hermitian systems are usually found from the bi-orthogonal set formed by the Bloch eigenstates of the Hamiltonian and by the eigenstates of the adjoint operator [14, 25]. Other non-standard topological invariants and non-Bloch Chern numbers have also been put forward recently [13, 16-18].

So far, most of the works in the literature deal with idealized and abstract Hamiltonians (e.g., extensions of the 1D Su-Schrieffer-Heeger model or the 2D Rice-Mele model, or others [13, 16-22]), that do not always connect in a straightforward way to realistic physical structures. In contrast, here we show that the topological phases of standard dispersive and lossy photonic materials can be characterized using Green function methods. The proposed theory generalizes the results of our earlier article [26] to non-Hermitian photonic systems. Even though we shall focus on optical materials, our analysis applies to both fermionic and bosonic platforms. While it is well-known that the Chern invariant of fermionic systems can be found from the system Green function [27-30], to our best knowledge, so far the application of this technique to non-Hermitian systems remains unexplored.

Our theory enables the calculation of the Chern invariants of non-Hermitian systems without relying on a gauge-dependent bi-orthogonal set of eigenstates, and thus may be advantageous and simpler from a computational point of view. More importantly, the Green function approach sheds light over several unsettled problems in the theory of topological non-Hermitian systems. In particular, it makes clear that the band-gaps correspond to strips of the complex plane wherein the system Green function is analytic (with no poles). Furthermore, by extending the arguments of Ref. [31], we show that for periodic systems described by linear differential-equations the spectrum must become gapless when the system is enclosed with "opaque-type" boundaries (often referred to as "open" in the literature). Thus, our theory settles in part the controversy about the application of the bulk-edge correspondence to non-Hermitian systems [10, 15, 16, 17, 19], and uncovers that a topological non-Hermitian cavity terminated with opaque-type walls must support edge states in the bulk band-gaps.

The article is organized as follows. In Sect. II, we present a motivation example that illustrates the application of the developed concepts to a magnetized electric plasma. In Sect. III we develop the general theory that enables finding the topological invariants of non-Hermitian systems from the system Green function. Then, in Sect. IV the theory is applied to photonic systems, i.e., systems whose dynamics is described by the Maxwell equations. In Sect. V, we generalize some of the ideas of Ref. [31] to non-Hermitian platforms and demonstrate that the Chern number integral depends critically on the Green function boundary conditions. Based on this result, we prove that consistent with the standard bulk-edge correspondence, the spectrum of a topologically nontrivial phase terminated with "opaque-type" boundaries must be gapless. Finally, Sect. VI contains a brief summary of the main findings.

## II. Motivation Example

To illustrate the application of the Green function methods that will be developed later in the article, next we compute the topological invariants of a lossy magnetized electric plasma (e. g., a semiconductor biased with a static magnetic field [32, 33]). It is assumed that the material is non-magnetic and that the relative permittivity tensor is of the form:

$$\overline{\overline{\varepsilon}} = \varepsilon_t \mathbf{1}_t + i\varepsilon_g \hat{\mathbf{z}} \times \mathbf{1}_t + \varepsilon_a \hat{\mathbf{z}} \otimes \hat{\mathbf{z}}, \tag{1}$$

with $\mathbf{1}_t = \hat{\mathbf{x}} \otimes \hat{\mathbf{x}} + \hat{\mathbf{y}} \otimes \hat{\mathbf{y}}$. The permittivity elements are ($\varepsilon_{11} = \varepsilon_{22} = \varepsilon_t$, $\varepsilon_{33} = \varepsilon_a$ and $\varepsilon_{12} = -\varepsilon_{21} = -i\varepsilon_g$):

$$\varepsilon_t = 1 - \frac{\omega_p^2 (1 + i\Gamma/\omega)}{(\omega + i\Gamma)^2 - \omega_0^2}, \qquad \varepsilon_g = \frac{1}{\omega} \frac{\omega_p^2 \omega_0}{\omega_0^2 - (\omega + i\Gamma)^2}, \qquad \varepsilon_a = 1 - \frac{\omega_p^2}{\omega(\omega + i\Gamma)}. \tag{2}$$

Here, $\omega_0 = -qB_0/m$ is the cyclotron frequency determined by the bias magnetic field $\mathbf{B}_0 = B_0 \hat{\mathbf{z}}$, $\Gamma$ is the collision frequency, $q = -e$ is the negative charge of the electrons, $m$ is the effective mass, and $\omega_p$ is the plasma frequency [34]. We focus on transverse magnetic (TM) waves with $\mathbf{H} = H_z \hat{\mathbf{z}}$, $\mathbf{E} = E_x \hat{\mathbf{x}} + E_y \hat{\mathbf{y}}$ and $\partial/\partial z = 0$. The electrodynamics is non-Hermitian when the collision frequency is nonzero $\Gamma > 0$. In this case the energy is not conserved as a function of time because of the material absorption, and hence the natural modes must have a finite lifetime, i.e., are associated with complex valued frequencies $\omega = \omega' + i\omega''$ and $\omega'' < 0$ (the mode lifetime is $\tau_{lf} = 1/(-2\omega'')$ [35]).

As extensively discussed in Refs. [36, 37], the characterization of the topological phases of a generic electromagnetic continuum requires the introduction of a high-frequency spatial cut-off in the material response. The spatial cut-off needs to ensure that the material response is asymptotically (in the limit $k \to \infty$) analogous to that of a reciprocal and non-bianisotropic dielectric. For a magnetized plasma the permittivity dispersion may be modified as [26, 36, 37]:

$$\varepsilon_t(k,\omega) = 1 + \frac{1}{1 + k^2/k_{max}^2}\left[\varepsilon_{t,\text{loc}}(\omega) - 1\right], \qquad \varepsilon_g(k,\omega) = \frac{\varepsilon_{g,\text{loc}}(\omega)}{1 + k^2/k_{max}^2}, \qquad (3)$$

where $k_{max}$ is the spatial cut-off. In the above, $\varepsilon_{t,\text{loc}}, \varepsilon_{g,\text{loc}}$ stand for the permittivity elements of a material with no spatial cut-off and are defined as in Eq. (2). The element $\varepsilon_a$ is irrelevant for TM-polarized waves and so it is not discussed here.

The dispersion of the TM-polarized bulk modes (plane waves propagating in the *xoy* plane with propagation factor $e^{ik_x x} e^{ik_y y}$) is given by [26, 36]:

$$k^2 = \varepsilon_{ef} \omega^2 / c^2, \qquad \varepsilon_{ef} = \left(\varepsilon_t^2 - \varepsilon_g^2\right)/\varepsilon_t, \qquad (4)$$

with $k^2 = k_x^2 + k_y^2$. It will be seen later that the Green function singularities in the complex frequency plane are determined by the solutions (with respect to $\omega$) of Eq. (4) with $k_x, k_y$ *real-valued*. Thus, the spectrum of the gyrotropic material is determined by the plane waves (Bloch modes) with $k_x, k_y$ real-valued.

Figure 1 represents the locus of the natural frequencies $\omega = \omega' + i\omega''$ in the complex plane (showing both the positive and negative frequency branches) associated with a real-valued wave vector for different values of the collision frequency $\Gamma$. Since the material response is invariant to rotations around the *z*-axis, it is evident that $\omega = \omega(k)$ with $0 \le k < \infty$. Therefore, in this example, the locus of $\omega = \omega' + i\omega''$ is typically formed by 5 disconnected curves, i.e., 5 disconnected *bands*. As seen, the bands are separated by vertical strips in the complex plane where the material does not support Bloch waves. These vertical strips correspond to the photonic band-gaps of the non-Hermitian material. Curiously, the band-gaps are nearly (but not exactly) independent on the value of $\Gamma$ in the range $0 \le \Gamma \le 0.5\omega_p$. Evidently, when the material is lossless ($\Gamma = 0$) the natural modes lie on the real-frequency axis. Note that when $\Gamma > 0$ there is a band of modes with frequencies along the imaginary frequency axis (black lines in Fig. 1). For sufficiently large values of $\Gamma$ the band-gaps close (not shown).

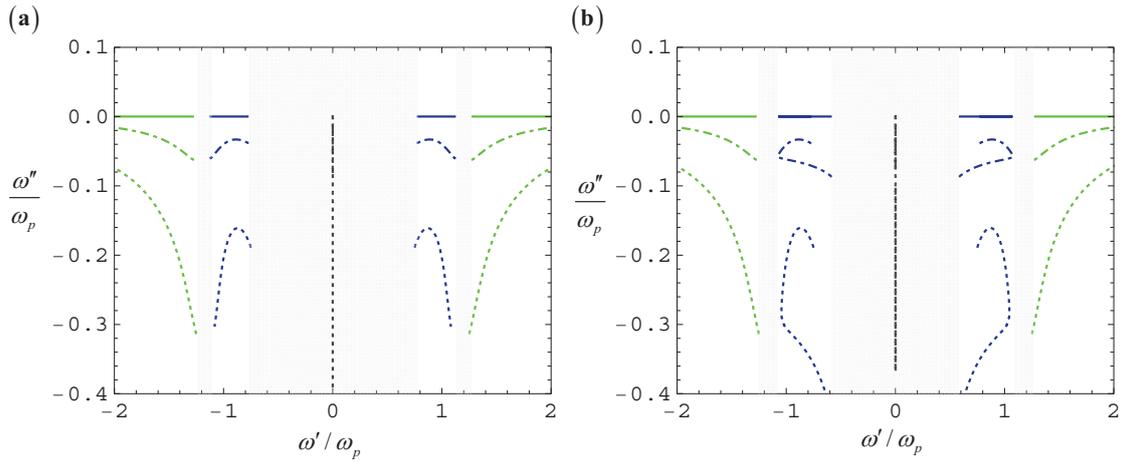

Fig. 1 Locus of the natural frequencies of the gyrotropic material with $\omega_0 = 0.5\omega_p$ in the complex-frequency plane ($\omega = \omega' + i\omega''$) for a real-valued wave vector. Solid lines: $\Gamma = 0$ (lossless case). Dot-dashed lines: $\Gamma = 0.1\omega_p$. Dotted lines: $\Gamma = 0.5\omega_p$. The shaded gray vertical strips represent the band-gaps in the complex plane. The left panel (a) shows the root locus for a local material and the right panel (b) for a material with a high-frequency cut-off $k_{max} = 10\omega_p / c$.

It is interesting to compare the spectrum of a local material with no high-frequency spatial cut-off (Fig. 1a) with the spectrum of a nonlocal material with $k_{max} = 10\omega_p / c$ (Fig. 1b). The two spectra are almost coincident for the natural modes with a small $k$, but may differ somewhat for eigenmodes with $k > k_{max}$ (compare the lower end of the blue curves in Figs. 1a and 1b).

Figure 2 depicts the explicit dependence of both $\omega'$ (panel a) and $\omega''$ (panel b) on the real-valued wave vector for a generic propagation direction in the xoy-plane. In particular, Fig. 2a shows that the real-part of the eigenfrequencies, $\omega' = \omega'(k_x)$, is rather insensitive to the value of the damping parameter $\Gamma$.

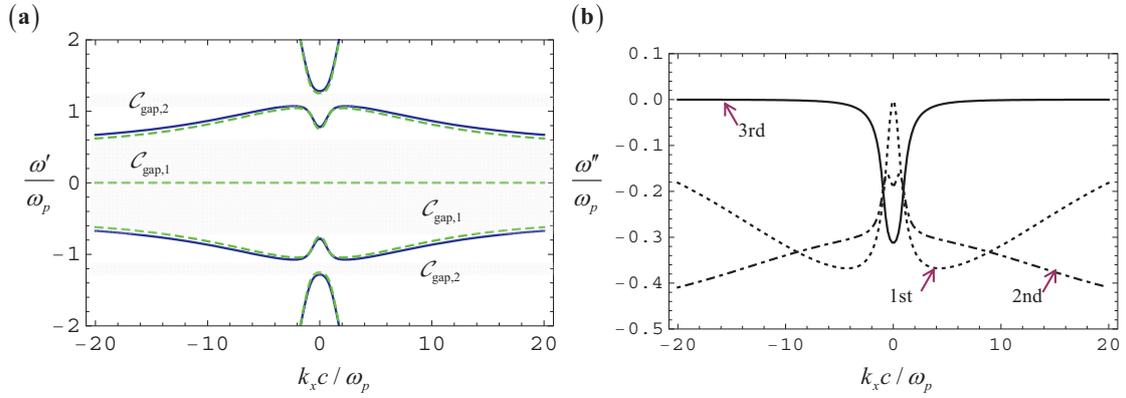

Fig. 2 Complex band structure of a gyrotropic material with $\omega_0 = 0.5\omega_p$ and $k_{max} = 10\omega_p/c$. (a) Real part of the natural oscillation frequencies as a function of $k_x$. Solid blue lines: $\Gamma = 0$ (lossless case). Dashed green lines: $\Gamma = 0.5\omega_p$. The shaded gray horizontal strips represent the band-gaps. (b) Imaginary parts of the natural frequencies of the 1st, 2nd and 3rd bands with non-negative frequency bands (ordered such that $0 = \omega'_1 < \omega'_2 < \omega'_3$) as a function of $k_x$ for $\Gamma = 0.5\omega_p$.

Similar to our previous article [26], the Chern invariants of the non-Hermitian material can be found by integrating the photonic Green function along a vertical line $\omega' = \text{Re}\{\omega\} = \omega_{gap}$ parallel to the imaginary frequency axis. The integration path $\text{Re}\{\omega\} = \omega_{gap}$ must be contained in the relevant band-gap, i.e., in one of the shaded gray regions in Fig. 1b. Moreover, the analysis of the following sections shows that for electric gyrotropic media (with the permittivity tensor (1)) with a high-frequency spatial cut-off $k_{max}$ the gap Chern number can be explicitly written as $\mathcal{C} = \mathcal{C}_1 + \mathcal{C}_2$ with:

$$\mathcal{C}_1 = \frac{-1}{\pi} \int_{-\infty}^{\infty} d\xi \int_0^{\infty} dk \, k \left(1 + \frac{2k^2}{k^2 + k_{max}^2}\right) \frac{\varepsilon_g}{\varepsilon_t} \tilde{\Phi} \left[ k^2 \partial_\omega \tilde{\Phi} + \frac{\omega}{c^2} \partial_\omega \left( \omega \varepsilon_{ef} \tilde{\Phi} \right) \right]_{\omega = \omega_{gap} + i\xi}. \quad (5a)$$

$$\mathcal{C}_2 = \frac{-1}{\pi} \int_0^{\infty} dk \int_{-\infty}^{\infty} d\xi \, \frac{k^3 \omega}{k^2 + k_{max}^2} \left\{ \frac{1}{\varepsilon_{ef}^2 \omega^2} \left(\frac{\varepsilon_g}{\varepsilon_t}\right)^2 \left( \omega \varepsilon_{ef} \tilde{\Phi} \partial_\omega \left(\frac{\varepsilon_g}{\varepsilon_t}\right) - \frac{\varepsilon_g}{\varepsilon_t} \partial_\omega \left( \omega \varepsilon_{ef} \tilde{\Phi} \right) \right) \right.$$

$$\left. + \frac{1}{c^2} \tilde{\Phi}^2 \left[ 2 \frac{\varepsilon_g}{\varepsilon_t} + \left(\frac{\varepsilon_g}{\varepsilon_t}\right)^2 \partial_\omega \left( \omega \frac{\varepsilon_g}{\varepsilon_t} \right) + \omega \varepsilon_{ef} \partial_\omega \left(\frac{\varepsilon_g}{\varepsilon_t}\right) - \frac{\varepsilon_g}{\varepsilon_t} \partial_\omega \left( \omega \varepsilon_{ef} \right) + \frac{c^2}{\varepsilon_{ef}^2 \omega^2} k^2 \partial_\omega \left( \omega \varepsilon_{ef} \frac{\varepsilon_g}{\varepsilon_t} \right) \right] \right\}_{\omega = \omega_{gap} + i\xi}$$

(5b)

where $\tilde{\Phi} = 1 / \left[ k^2 - (\omega/c)^2 \varepsilon_{ef}(k, \omega) \right]$, $\varepsilon_{ef} = \left( \varepsilon_t^2 - \varepsilon_g^2 \right) / \varepsilon_t$, and $\varepsilon_t, \varepsilon_g$ are defined as in Eq. (3). Equation (5) coincides exactly with Eq. (43) of Ref. [26], which was originally derived for lossless systems, but remains valid in the non-Hermitian case. The justification of this property will be given in Sect. IV. Note that the integrand of Eq. (5) is singular in the complex plane when $\tilde{\Phi}$ is singular, i.e., when $\omega$ satisfies Eq. (4) for some real-valued wave vector (Bloch eigenfrequency).

Using Eq. (5) the gap Chern numbers of the gyrotropic material were numerically found as a function of the collision frequency $\Gamma$ (see Fig. 3). The gap Chern numbers are labeled as in Fig. 2a. Due to the particle hole-symmetry of Maxwell's equations, the gap Chern numbers of the negative frequency gaps are identical to those of the corresponding positive frequency gaps. Note that the gap Chern numbers take into account the contributions of all bands below the gap (modes with $\text{Re}\{\omega\} < \omega_{gap}$), including the negative frequency bands [26]. As seen, the gap Chern numbers are independent of the value of $\Gamma$ because the complex band-gap remains open, i.e., the bands remain separated by a vertical strip in the complex plane when loss is introduced into the material response. In particular, it follows that nonreciprocal photonic materials can be topologically nontrivial even in presence of rather strong material absorption. The band-gaps associated with $\mathcal{C}_{\text{gap},2}$ close for a collision frequency $\Gamma$ slightly larger than $\omega_p$.

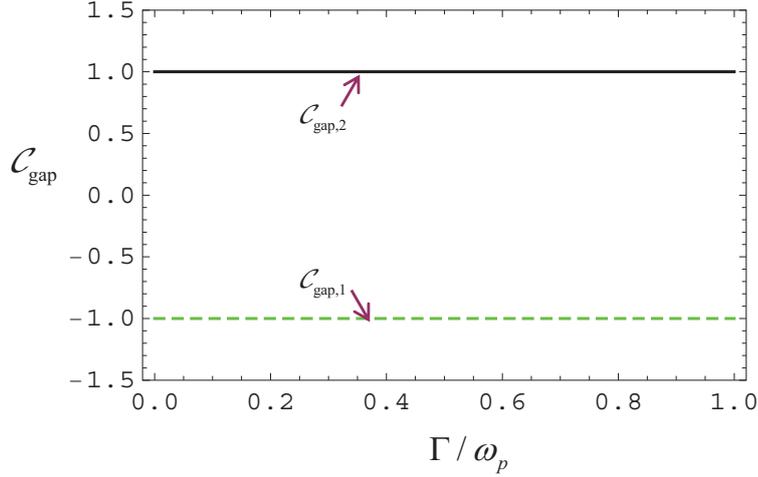

Fig. 3 Gap Chern numbers of a gyrotropic material with $\omega_0 = 0.5\omega_p$ and $k_{\max} = 10\omega_p/c$ as a function of $\Gamma$. The gap Chern numbers are labeled as in Fig. 2a.

## III. General Theory

In the following, we develop a general Green function formalism to topologically classify the phases of non-Hermitian (fermionic or bosonic) systems.

### A. Complex band structure

We consider generic operators $\hat{L}_\mathbf{k}$ and $\mathbf{M}_g$ that may be non-Hermitian. The operator $\hat{L}_\mathbf{k}$ is parameterized by the *real-valued* wave vector $\mathbf{k} = k_x\hat{\mathbf{x}} + k_y\hat{\mathbf{y}}$. For clarity, for now the family of operators $\hat{L}_\mathbf{k}$ is assumed periodic in $\mathbf{k}$ with the irreducible cell (typically a Brillouin zone) denoted by BZ. For example, in lattice-type models of physical systems (e.g., tight-binding models) the space is discretized into a grid of points and $\hat{L}_\mathbf{k}$ is a matrix periodic in $\mathbf{k}$. In general, the periodicity in $\mathbf{k}$ is not mandatory and more relaxed conditions may be enforced (these will be made precise later). The operator $\mathbf{M}_g$ is independent of $\mathbf{k}$ and invertible.

We introduce the following generalized eigenvalue problem:

$$\hat{L}_\mathbf{k} \cdot \mathbf{Q}_{n\mathbf{k}} = \omega_{n\mathbf{k}} \mathbf{M}_g \cdot \mathbf{Q}_{n\mathbf{k}} \quad (n=1,2,\ldots). \tag{6}$$

Here, $\mathbf{Q}_{n\mathbf{k}}$ are the generalized eigenstates of $\hat{L}_\mathbf{k}$ and $\omega_{n\mathbf{k}}$ are the generalized eigenvalues. If $\mathbf{M}_g$ is taken as the identity operator, Eq. (6) becomes a standard eigenvalue problem. The operators $\hat{L}_\mathbf{k}$ and $\mathbf{M}_g$ do not need to be Hermitian, and thereby the $\omega_{n\mathbf{k}}$ are generally complex-valued. Thus, $\mathbf{Q}_{n\mathbf{k}}$ and $\omega_{n\mathbf{k}}$ determine the "complex band structure" of $\hat{L}_\mathbf{k}$.

Let us suppose that $\hat{L}_\mathbf{k}$ has no eigenvalues in some vertical strip, $\omega_L < \mathrm{Re}\{\omega\} < \omega_U$, of the complex-frequency plane, analogous to the example of Fig. 1. Then, we say that the operator $\hat{L}_\mathbf{k}$ has a complete band-gap in the relevant strip. The band-gap separates the "bands" into two classes: (i) those formed by eigenvectors with $\mathrm{Re}\{\omega_{n\mathbf{k}}\} < \omega_{\mathrm{gap}}$, which we shall refer to as the filled ($F$) bands, and (ii) those formed by eigenvectors with $\mathrm{Re}\{\omega_{n\mathbf{k}}\} > \omega_{\mathrm{gap}}$, which we refer to as the empty ($E$) bands. Here, $\omega_{\mathrm{gap}}$ is a generic (real-valued) frequency in the gap ($\omega_L < \omega_{\mathrm{gap}} < \omega_U$).

## B. Chern topological number

We introduce a Green function operator defined by:

-10-

$$\mathcal{G}_{\mathbf{k}}(\omega) = i\left(\hat{L}_{\mathbf{k}} - \mathbf{M}_g \omega\right)^{-1}. \tag{7}$$

Evidently, the Green function operator has poles at the eigenfrequencies $\omega = \omega_{n\mathbf{k}}$, but otherwise is an analytic function of frequency. In particular, it is analytic over the vertical strip of the complex plane that determines a complete band-gap ($\omega_L < \text{Re}\{\omega\} < \omega_U$). The inverse operator is such that $i\mathcal{G}_{\mathbf{k}}^{-1} = \hat{L}_{\mathbf{k}} - \mathbf{M}_g \omega$. For convenience, we denote

$$\partial_j \mathcal{G}_{\mathbf{k}}^{-1} = \frac{\partial}{\partial k_j} \mathcal{G}_{\mathbf{k}}^{-1} = -i\frac{\partial \hat{L}_{\mathbf{k}}}{\partial k_j} \quad (j=1,2) \text{ with } k_1 = k_x \text{ and } k_2 = k_y.$$ Notice that $\partial_j \mathcal{G}_{\mathbf{k}}^{-1}$ is independent of frequency. The Chern topological number associated with the band gap is defined by means of the Green function operator as:

$$\mathcal{C} = \frac{1}{(2\pi)^2} \iint_{B.Z.} d^2\mathbf{k} \int_{\omega_{\text{gap}} - i\infty}^{\omega_{\text{gap}} + i\infty} d\omega \, \text{Tr}\left\{ \partial_1 \mathcal{G}_{\mathbf{k}}^{-1} \cdot \mathcal{G}_{\mathbf{k}} \cdot \partial_2 \mathcal{G}_{\mathbf{k}}^{-1} \cdot \partial_\omega \mathcal{G}_{\mathbf{k}} \right\}, \tag{8}$$

where $\partial_\omega = \partial/\partial\omega$. The integral in $\omega$ is over the line $\text{Re}\{\omega\} = \omega_{\text{gap}}$ parallel to the complex-frequency imaginary axis and $\text{Tr}\{...\}$ stands for the trace operator. In our previous article [26], it was shown that the photonic gap Chern number can be expressed as in Eq. (8) when the relevant system is formed by lossless materials, i.e., when the operators are Hermitian (the link between the theory of Ref. [26] and Eq. (8) is further discussed in Sect. IV). In the following, it is shown that for non-Hermitian operators $\mathcal{C}$ remains a topological integer.

Specifically, let us suppose that the non-Hermitian operator of interest may be regarded as a smooth deformation of some Hermitian operator. For example, consider that $\hat{L}_{\mathbf{k}} \equiv \hat{L}_{\mathbf{k}}(1)$ is the non-Hermitian operator of interest, with $\hat{L}_{\mathbf{k}}(\alpha)$ a smooth function of the parameter $\alpha$ ($0 \leq \alpha \leq 1$), such that $\hat{L}_{\mathbf{k}}(0)$ is Hermitian. In photonic systems the parameter $\alpha$ is typically related to a damping factor, e.g., a collision frequency, analogous to the example of Sect. II. Thereby, the Green function operator $\mathcal{G}_\mathbf{k} = \mathcal{G}_\mathbf{k}(\alpha)$ and the Chern number $\mathcal{C} = \mathcal{C}(\alpha)$ are functions of $\alpha$. Inspired by Ref. [27], we prove in Appendix A that provided the deformation does not close the relevant band-gap (so that $\omega_{gap}$ in Eq. (8) is fixed), then the derivative of $\mathcal{C}$ with respect to the parameter $\alpha$ is given by:

$$\frac{\partial \mathcal{C}}{\partial \alpha} = \frac{1}{(2\pi)^2} \frac{1}{2} \iint_{B.Z.} d^2\mathbf{k} \int_{\omega_{gap}-i\infty}^{\omega_{gap}+i\infty} d\omega\, \varepsilon^{ijl} \partial_i \mathrm{Tr}\left\{ \partial_\alpha \mathcal{G}_\mathbf{k} \cdot \partial_j \mathcal{G}_\mathbf{k}^{-1} \cdot \mathcal{G}_\mathbf{k} \cdot \partial_l \mathcal{G}_\mathbf{k}^{-1} \right\}. \tag{9}$$

In the above, $\varepsilon^{ijl}$ is the Levi-Civita symbol, $\partial_0 = \partial/\partial\omega$, and the summation over all $i,j,l \in \{0,1,2\}$ is implicit. Notice that the integrand is a sum of derivatives of the trace operator. Thus, if the term $\mathrm{Tr}\{...\}$ satisfies suitable conditions at the boundary of the BZ and at $\omega = \infty$ the right-hand side of Eq. (9) vanishes.

A system is topological when it is possible to guarantee that the integral in Eq. (9) vanishes. The periodicity of $\hat{L}_\mathbf{k}$ in $\mathbf{k}$ is a sufficient (but not a necessary) condition to ensure that. Indeed, if $\hat{L}_\mathbf{k}$ is a periodic function of $\mathbf{k}$ then $\mathcal{G}_\mathbf{k}$ also is. In these conditions, the contribution from the boundary of BZ vanishes, and therefore it follows that $\frac{\partial \mathcal{C}}{\partial \alpha} = 0$. But since $\mathcal{C}_{\alpha=0}$ is an integer (because by hypothesis $\hat{L}_\mathbf{k}(0)$ is Hermitian) it follows that $\mathcal{C} = \mathcal{C}_{\alpha=1}$ also is. Therefore, the gap Chern number is an integer totally insensitive to any possible non-Hermitian deformation that does not close the band-gap; this property unveils the topological nature of non-Hermitian platforms.

In summary, if the operators $\hat{L}_\mathbf{k}$ guarantee that the integral in Eq. (9) vanishes (e.g., if $\hat{L}_\mathbf{k}$ is periodic) and if $\hat{L}_\mathbf{k}$ may be regarded as a deformation of some Hermitian system then $\mathcal{C}$ is a topological integer.

## C. Berry potential and Berry curvature

The approach of the previous subsection is rather powerful and relies on the well established topological properties of Hermitian systems. It is instructive to prove that the gap Chern number is an integer following a different path, namely by diagonalizing the Green function operator. This second method is related to the theory of Ref. [14] and enables us to introduce in a natural manner the notions of Berry potential and curvature for non-Hermitian systems.

To begin with, it is convenient to define $\hat{\mathcal{L}}_\mathbf{k} = \mathbf{M}_g^{-1/2} \hat{L}_\mathbf{k} \mathbf{M}_g^{-1/2}$. From Eq. (6) it is seen that $\hat{\mathcal{L}}_\mathbf{k} \cdot \mathcal{Q}_{n\mathbf{k}} = \omega_{n\mathbf{k}} \mathcal{Q}_{n\mathbf{k}}$ with $\mathcal{Q}_{n\mathbf{k}} = \mathbf{M}_g^{1/2} \cdot \mathbf{Q}_{n\mathbf{k}}$ the eigenvectors of $\hat{\mathcal{L}}_\mathbf{k}$. It will be assumed throughout that the $\mathcal{Q}_{n\mathbf{k}}$ are ordered in such a way that the filled bands correspond to the indices $n = 1, 2, ... N_F$, and the empty bands to the indices $n = N_F + 1, ....$, with $N_F$ the number of filled bands.

When $\hat{\mathcal{L}}_\mathbf{k}$ is Hermitian the spectral theorem guarantees that its eigenfunctions $\mathcal{Q}_{n\mathbf{k}}$ form a complete set of the relevant vector space. However, for general non-Hermitian operators, e.g., when the operator has exceptional points [10, 38], this property does not necessarily hold. Here, we focus on the class of "diagonalizable" operators $\hat{\mathcal{L}}_\mathbf{k}$ whose eigenfunctions span the entire space, even when $\hat{\mathcal{L}}_\mathbf{k}$ is non-Hermitian. For example, a $\hat{\mathcal{L}}_\mathbf{k}$ that is a sufficiently weak perturbation of a Hermitian operator must have that property.

The operator $\hat{\mathcal{L}}_\mathbf{k}$ can thus be represented by a diagonal matrix $\Omega_\mathbf{k} = \left[\omega_{n\mathbf{k}}\delta_{m,n}\right]_{m,n=1,2,...}$ with respect to the basis determined by $\mathcal{Q}_{n\mathbf{k}}$. It is convenient to introduce a fixed (but otherwise arbitrary) basis of the relevant vector space, $\mathbf{e}_1, \mathbf{e}_2,...$, with elements independent of the wave vector. Let $S_\mathbf{k}$ be the matrix with the coordinates of $\mathcal{Q}_{n\mathbf{k}}$ in the basis $\mathbf{e}_1, \mathbf{e}_2,...$ (specifically, the first column of $S_\mathbf{k}$ has the coordinates of $\mathcal{Q}_{1\mathbf{k}}$ in the $\mathbf{e}_n$ basis, etc). Then, $\hat{\mathcal{L}}_\mathbf{k}$ is represented by the matrix $S_\mathbf{k} \cdot \Omega_\mathbf{k} \cdot S_\mathbf{k}^{-1}$ in the $\mathbf{e}_n$ basis. Evidently, the matrix $S_\mathbf{k}$ depends on the considered eigenfunctions $\mathcal{Q}_{n\mathbf{k}}$, which are normalized arbitrarily (both in amplitude and phase). Thus, $S_\mathbf{k}$ is gauge dependent.

Using the proposed matrix representation of $\hat{\mathcal{L}}_\mathbf{k}$ it is proven in Appendix B that $\mathcal{C}$ can be written as:

$$\mathcal{C} = \frac{1}{2\pi} \iint_{B.Z.} d^2\mathbf{k}\, \hat{\mathbf{z}} \cdot (\nabla_\mathbf{k} \times \mathcal{A}_\mathbf{k}) = \frac{1}{2\pi} \iint_{B.Z.} d^2\mathbf{k}\, (\partial_1 \mathcal{A}_2 - \partial_2 \mathcal{A}_1), \tag{10}$$

where by definition $\mathcal{A}_\mathbf{k} = i\,\mathrm{Tr}\{S_\mathbf{k}^{-1} \cdot \partial_\mathbf{k} S_\mathbf{k} \cdot \mathbf{1}_F\}$ is the Berry potential. Here, $\mathbf{1}_F = \sum_{\mathrm{Re}\{\omega_{n\mathbf{k}}\}<\omega_{\mathrm{gap}}} \hat{\mathbf{u}}_n \otimes \hat{\mathbf{u}}_n$ represents a diagonal matrix (independent of the wave vector) with diagonal elements identical to "1" for $n = 1,2,...N_F$ (the filled bands) and equal to zero otherwise. It should be noted that $\mathcal{A}_\mathbf{k}$ is generally complex-valued. We will see below that $\mathcal{C}$ given by Eq. (10) is necessarily a real-number, and hence it is possible to take $\mathcal{A}_\mathbf{k} = \mathrm{Re}\{i\,\mathrm{Tr}\{S_\mathbf{k}^{-1} \cdot \partial_\mathbf{k} S_\mathbf{k} \cdot \mathbf{1}_F\}\}$ without affecting the value of $\mathcal{C}$. It can be shown that the latter definition of $\mathcal{A}_\mathbf{k}$ is consistent with that of Ref. [14].

Clearly, when $S_\mathbf{k}$ is globally defined as a smooth periodic function over *BZ* the Stokes theorem implies that the gap Chern number vanishes. Hence, similar to the Hermitian case [27, 36], a nontrivial $\mathcal{C}$ indicates the impossibility of choosing a globally defined smooth basis of eigenfunctions $\mathcal{Q}_{n\mathbf{k}}$, and implies that there is an obstruction to the application of the Stokes theorem to the entire wave vector domain.

As usual, the Berry potential is gauge dependent. For example, a gauge transformation of the form $\mathcal{Q}_{n\mathbf{k}} \to \alpha_{n\mathbf{k}} e^{i\theta_{n\mathbf{k}}} \mathcal{Q}_{n\mathbf{k}}$, with $\alpha_{n\mathbf{k}} > 0$ an amplitude factor and $\theta_{n\mathbf{k}}$ a phase factor, leads to a different $S_\mathbf{k}$, and thereby to a different Berry potential. It is shown in Appendix A, that for two generic bases of eigenfunctions $\mathcal{Q}_{n\mathbf{k}}$ and $\mathcal{Q}'_{n\mathbf{k}}$ the corresponding Berry potentials are linked by:

$$\mathcal{A}'_\mathbf{k} = \mathcal{A}_\mathbf{k} + \partial_\mathbf{k}(i \ln g_\mathbf{k}), \tag{11}$$

with $g_\mathbf{k}$ some smooth (single-valued) function of $\mathbf{k}$ (in the domain wherein both $\mathcal{Q}_{n\mathbf{k}}$ and $\mathcal{Q}'_{n\mathbf{k}}$ are smooth). In particular, it follows that the Berry curvature $\mathcal{F}_\mathbf{k} = \hat{\mathbf{z}} \cdot (\nabla_\mathbf{k} \times \mathcal{A}_\mathbf{k})$ is gauge invariant also for non-Hermitian operators.

In order to demonstrate that $\mathcal{C}$ is really an integer, we mimic the arguments of Ref. [36]. For simplicity, it is supposed that there is some globally defined smooth basis of eigenvectors, $\mathcal{Q}_{n\mathbf{k}}$, except at a finite number of singular points ($\mathbf{k}_{S,i}$) in the BZ. Then, using Stokes theorem in Eq. (10), it follows that the gap Chern number can be written as a line integral of the Berry potential around the singular points: $\mathcal{C} = -\frac{1}{2\pi}\sum_i \oint_{C_i} \mathcal{A}_\mathbf{k} \cdot d\mathbf{l}$.

Here, $C_i$ is a circle of infinitesimal radius centered at $\mathbf{k}_{S,i}$. The contribution of each singular point is necessarily an integer. The reason is that it is possible to pick another gauge $\mathcal{Q}'_{n\mathbf{k}}$ that is smooth in the vicinity of $\mathbf{k}_{S,i}$. The corresponding Berry potential is linked to $\mathcal{A}_\mathbf{k}$ as in Eq. (11). The line contour of $\mathcal{A}'_\mathbf{k}$ around circle of infinitesimal radius vanishes because $\mathcal{A}'_\mathbf{k}$ is a smooth function. Therefore, $\oint_{C_i} \mathcal{A}_\mathbf{k} \cdot \mathbf{dl} = \oint_{C_i} \partial_\mathbf{k} (i \ln g_\mathbf{k}) \cdot \mathbf{dl} = 2\pi n$, for some integer $n$. The latter identity follows from the fact that the logarithm is a multi-valued function with the different branches differing by $i2\pi n$ (note that $g_\mathbf{k}$ is continuous over $C_i$ but $\ln g_\mathbf{k}$ may not be). This confirms that $\mathcal{C}$ given by Eq. (8) is really a topological integer for non-Hermitian systems.

The previous analysis assumes that the Hamiltonian is diagonalizable. However, the conclusion that $\mathcal{C}$ is an integer can be readily extended to non-diagonalizable Hamiltonians that are smooth deformations ($\hat{L}_\mathbf{k} = \hat{L}_\mathbf{k}(\alpha)$) of any diagonalizable (possibly non-Hermitian) Hamiltonian. Indeed, provided the band-gap is not closed by the transformation $\hat{L}_\mathbf{k}(\alpha)$ the theory of Sect. II.B implies that $\partial \mathcal{C} / \partial \alpha = 0$. Thus, the developed theory guarantees that for a wide class of Hamiltonians with exceptional points [10, 38] $\mathcal{C}$ given by Eq. (8) must be an integer number.

## D. Systems described by linear differential-equations

Let us now focus on problems where $\hat{L}_\mathbf{k}$ and $\mathbf{M}_g$ are differential operators of the form $\hat{L}_\mathbf{k} = \hat{L}_\mathbf{k}(\mathbf{r}, -i\nabla)$ and $\mathbf{M}_g = \mathbf{M}_g(\mathbf{r})$ that act on functions defined over some volumetric region of interest (denoted in the following by "cell"). In this context, it is convenient to trace out the degrees of freedom associated with the spatial coordinates. This can be done considering a basis of kets $|\mathbf{r}\rangle$ normalized such that $\langle \mathbf{r} | \mathbf{r}' \rangle = \delta(\mathbf{r} - \mathbf{r}')$, so that $\mathbf{1} = \int d^3\mathbf{r} \, |\mathbf{r}\rangle\langle\mathbf{r}|$. Then, the trace operator is of the form $\mathrm{Tr}\{...\} = \int d^3\mathbf{r} \, \mathrm{tr}\{\langle\mathbf{r}|...|\mathbf{r}\rangle\}$ where $\mathrm{tr}\{...\}$ is the trace over degrees of freedom unrelated to the spatial coordinates (e.g., associated with the polarization of the electromagnetic field). Introducing $\mathcal{G}_\mathbf{k}(\mathbf{r},\mathbf{r}',\omega) = \langle\mathbf{r}|\mathcal{G}_\mathbf{k}|\mathbf{r}'\rangle$ it can be shown that Eq. (8) reduces to:

$$\mathcal{C} = \frac{1}{(2\pi)^2} \iint_{BZ} d^2\mathbf{k} \int_{\omega_{\mathrm{gap}}-i\infty}^{\omega_{\mathrm{gap}}+i\infty} d\omega \int_{\mathrm{cell}}\int_{\mathrm{cell}} dV dV' \, \mathrm{tr}\left\{\left[\partial_2 \hat{L}_\mathbf{k} \cdot \mathcal{G}_\mathbf{k}(\mathbf{r},\mathbf{r}',\omega)\right] \cdot \left[\partial_1 \hat{L}_\mathbf{k} \cdot \partial_\omega \mathcal{G}_\mathbf{k}(\mathbf{r}',\mathbf{r},\omega)\right]\right\}. \quad (12)$$

To obtain the above formula we used $\langle\mathbf{r}|\partial_i \mathcal{G}_\mathbf{k}^{-1}|\mathbf{r}'\rangle = -i\delta(\mathbf{r}-\mathbf{r}')\partial_i \hat{L}_\mathbf{k}(\mathbf{r}',-i\nabla')$ and $\mathbf{1} = \int d^3\mathbf{r} \, |\mathbf{r}\rangle\langle\mathbf{r}|$. The gradient operator of $\partial_1 \hat{L}_\mathbf{k}$ acts exclusively on the $\mathbf{r}'$ coordinate of $\partial_\omega \mathcal{G}_\mathbf{k}(\mathbf{r}',\mathbf{r})$, whereas the gradient operator of $\partial_2 \hat{L}_\mathbf{k}$ acts on the $\mathbf{r}$ coordinate of $\mathcal{G}_\mathbf{k}(\mathbf{r},\mathbf{r}')$. From Eq. (7) one has $\left(\hat{L}_\mathbf{k} - \mathbf{M}_g \omega\right)\mathcal{G}_\mathbf{k} = i\mathbf{1}$ and thereby $\mathcal{G}_\mathbf{k}(\mathbf{r},\mathbf{r}',\omega)$ is the Green function of the system, i.e., the solution of:

$$\left(\hat{L}_\mathbf{k}(\mathbf{r},-i\nabla) - \mathbf{M}_g(\mathbf{r})\omega\right) \cdot \mathcal{G}_\mathbf{k}(\mathbf{r},\mathbf{r}',\omega) = i\mathbf{1}\delta(\mathbf{r}-\mathbf{r}'). \quad (13)$$

Typically, $\mathcal{G}_\mathbf{k}(\mathbf{r},\mathbf{r}',\omega)$ is a finite dimension matrix and thereby the $\mathrm{tr}\{...\}$ operator in Eq. (12) is the standard matrix trace operator.

## IV. Application to Photonic Systems

It is straightforward to apply the developed ideas to non-Hermitian photonic platforms. Our analysis is focused on lossy systems (all the materials are passive), but the formalism can be extended in a trivial manner to systems with gain elements (with active materials).

We follow the notations of our previous works [26, 36], so that the frequency-domain source-free Maxwell's equations can be written in the compact form as $\hat{N} \cdot \mathbf{f} = \omega \mathbf{M}(\mathbf{r}, \omega) \cdot \mathbf{f}$ with $\mathbf{f} = (\mathbf{E} \quad \mathbf{H})^T$ the electromagnetic field (represented by a six-vector), $\mathbf{M}(\mathbf{r}, \omega)$ is the material matrix, and $\hat{N}$ is a differential operator defined by

$$\hat{N}(-i\nabla) = \begin{pmatrix} \mathbf{0} & i\nabla \times \mathbf{1}_{3\times 3} \\ -i\nabla \times \mathbf{1}_{3\times 3} & \mathbf{0} \end{pmatrix}, \tag{14}$$

with $\mathbf{1}_{3\times 3}$ the identity matrix of dimension three. For example, for standard non-magnetic media the material matrix is of the form $\mathbf{M} = \begin{pmatrix} \varepsilon_0 \overline{\varepsilon} & \mathbf{0} \\ \mathbf{0} & \mu_0 \mathbf{1}_{3\times 3} \end{pmatrix}$. Without loss of generality, we focus on the case wherein all the poles of the material matrix lie either on the real axis or in the lower-half complex plane (passive material). Similar to the lossless case [26, 36, 39, 40, 41], it is shown in Appendix C that when the material matrix is a meromorphic function of frequency the electrodynamics of a dispersive lossy system can always be reduced to a standard Schrödinger-type time-evolution problem. In such a framework, the state vector is of the form $\mathbf{Q} = (\mathbf{f} \quad \mathbf{Q}^{(1)} \quad ...)^T$. The first component of $\mathbf{Q}$ determines the electromagnetic fields $\mathbf{f}$; the remaining components ($\mathbf{Q}^{(\alpha)}$) are related to the internal degrees of freedom of the materials. The eigenmodes of the generalized problem are the solutions of

$$\hat{L}(\mathbf{r}, -i\nabla) \cdot \mathbf{Q} = \omega \mathbf{M}_g(\mathbf{r}) \cdot \mathbf{Q}. \tag{15}$$

The operators $\hat{L}$ and $\mathbf{M}_g$ can be explicitly written as a function of the original material matrix, as detailed in the Appendix C. The topological classification of a photonic system is based on the generalized eigenvalue problem (15).

## A. Periodic systems

As a first example, we consider periodic (fully three-dimensional) waveguide-type photonic crystals, such that the material matrix is periodic in the coordinates $x$ and $y$: $\mathbf{M}(x+a_1,y,z) = \mathbf{M}(x,y,z) = \mathbf{M}(x,y+a_2,z)$. Furthermore, the system is assumed to be closed along the $z$-direction, e.g., it can be terminated with metallic plates placed at $z=0$ and $z=d$ (bottom and top walls, respectively), or with any other boundary conditions that force the waves to flow along directions parallel to the *xoy* plane. There is a common misconception that the Chern topological classification only applies to two-dimensional systems. This is not correct. Generally speaking, the space dimension can be arbitrary (greater or equal than two), but the wave propagation should be constrained to directions parallel to some plane (e.g., the *xoy* plane). In these conditions, the electromagnetic modes can be classified as Bloch waves labeled by a two-dimensional wave vector $\mathbf{k} = k_x \hat{\mathbf{x}} + k_y \hat{\mathbf{y}}$, even if the space dimension exceeds 2. For a 3D waveguide-type photonic crystal the Bloch waves correspond to waveguide modes that propagate in the region delimited by the top and bottom walls.

Let $\mathbf{Q}_{n\mathbf{k}}(\mathbf{r})$ be the envelope of a generic (Bloch) waveguide mode ($\mathbf{Q} = \mathbf{Q}_{n\mathbf{k}} e^{i\mathbf{k}\cdot\mathbf{r}}$) associated with the eigenfrequency $\omega_{n\mathbf{k}}$. From Eq. (15) it readily follows that $\mathbf{Q}_{n\mathbf{k}}$ satisfies the general Eq. (6) with

$$\hat{L}_{\mathbf{k}}(\mathbf{r},-i\nabla) \equiv \hat{L}(\mathbf{r},-i\nabla+\mathbf{k}). \tag{16}$$

Even though $\hat{L}_{\mathbf{k}}$ is not periodic (it is typically linear in $\mathbf{k}$) the topological classification remains feasible [42]. In particular, when the photonic system has a full photonic band gap in some vertical strip of the complex plane, the corresponding gap Chern number can be evaluated using Eq. (12) with the Green function $\mathcal{G}_\mathbf{k}(\mathbf{r},\mathbf{r}',\omega)$ defined as in Eq. (13) and BZ the photonic crystal Brillouin zone. The Green function $\mathcal{G}_\mathbf{k}(\mathbf{r},\mathbf{r}',\omega)$ satisfies periodic boundary conditions over (the lateral walls) of a unit cell. In a band-gap the non-Hermitian waveguide does not support any guided mode with a real-valued wave vector.

For media without spatial dispersion $\hat{L}_\mathbf{k}$ is linear in $\mathbf{k}$, and thereby the operators $\partial_i \hat{L}_\mathbf{k}$ are independent of the wave vector (and of the $\nabla$ operator). Furthermore, in such conditions $\partial_i \hat{L}_\mathbf{k}$ only acts on the electromagnetic degrees of freedom ($\mathbf{f}$). Therefore, the gap Chern number can be written in terms of the "electromagnetic component", $\overline{\mathbf{G}}_\mathbf{k}$, of the Green function $\mathcal{G}_\mathbf{k}(\mathbf{r},\mathbf{r}',\omega)$. Specifically, similar to our previous article [26], it can be shown that:

$$\mathcal{C} = \frac{1}{(2\pi)^2} \iint_{BZ} d^2\mathbf{k} \int_{\omega_{gap}-i\infty}^{\omega_{gap}+i\infty} d\omega \int_{cell}\int_{cell} dV dV' \, \text{tr}\left\{\partial_2 \hat{N} \cdot \overline{\mathbf{G}}_\mathbf{k}(\mathbf{r},\mathbf{r}',\omega) \cdot \partial_1 \hat{N} \cdot \partial_\omega \overline{\mathbf{G}}_\mathbf{k}(\mathbf{r}',\mathbf{r},\omega)\right\}. \quad (17)$$

In the above, $\partial_i \hat{N}$ ($i$=1,2) stands for the constant matrix $\begin{pmatrix} \mathbf{0} & -\hat{\mathbf{u}}_i \times \mathbf{1}_{3\times 3} \\ \hat{\mathbf{u}}_i \times \mathbf{1}_{3\times 3} & \mathbf{0} \end{pmatrix}$ with $\hat{\mathbf{u}}_1 = \hat{\mathbf{x}}$ and $\hat{\mathbf{u}}_2 = \hat{\mathbf{y}}$. The Green function $\overline{\mathbf{G}}_\mathbf{k}$ corresponds to the $6\times 6$ upper-block of $\mathcal{G}_\mathbf{k}$, and it can be shown that it satisfies:

$$\hat{N}(-i\nabla + \mathbf{k}) \cdot \overline{\mathbf{G}}_\mathbf{k}(\mathbf{r},\mathbf{r}',\omega) = \omega \mathbf{M}(\mathbf{r},\omega) \cdot \overline{\mathbf{G}}_\mathbf{k}(\mathbf{r},\mathbf{r}',\omega) + i\mathbf{1}\delta(\mathbf{r}-\mathbf{r}'), \quad (18)$$

subject to periodic boundary conditions in a unit cell. Importantly, $\overline{\mathbf{G}}_\mathbf{k}$ only depends on the electromagnetic degrees of freedom of the problem and on the dispersive material matrix. Thus, the gap Chern number can be found directly from the photonic Green function [26], even for non-Hermitian systems. In Appendix D, it is shown that Eq. (17) agrees with the result of Ref. [26].

In summary, the theory of Ref. [26] was extended to non-Hermitian systems. The final result is unmodified: the gap Chern number is expressed exactly by the same formulas as in the Hermitian case. Nevertheless, different from the case of Hermitian systems [26] the domain of integration in Eq. (17) (over a line parallel to the imaginary frequency axis) cannot be reduced to a semi-straight line in the upper-half frequency plane because the presence of loss breaks the mirror symmetry with respect to the real-frequency axis.

The gap Chern number $\mathcal{C}$ includes the contribution of all bands below the gap, including the negative frequency bands. For completeness, we note that the topological charge associated with only the positive frequency bands is given by

$$\mathcal{C}_+ = \frac{1}{(2\pi)^2} \iint_{BZ} d^2\mathbf{k} \int_{\text{cell}} \int_{\text{cell}} dV dV' \left[ \int_{\omega_{\text{gap}}-i\infty}^{\omega_{\text{gap}}+i\infty} d\omega - \int_{0-i\infty}^{0+i\infty} d\omega \right] \{...\},$$

where the integrand (i.e., the term inside brackets) is the same as in Eq. (17). Possible poles on the imaginary frequency axis should be avoided by deforming slightly the integration contour so that it is contained in the semi-plane $\text{Re}\{\omega\} > 0$. The total topological charge of the negative frequency bands may be nonzero [26]. This happens, for example, when the nonreciprocal response is so strong in the static limit that it leads to an exchange of topological charge between the positive and negative frequency bands [43, 44]. Thus, in general, the two Chern numbers are different $\mathcal{C} \neq \mathcal{C}_+$. The most relevant quantity is $\mathcal{C}$ not $\mathcal{C}_+$: the bulk-edge correspondence links the gap Chern number with the net number of unidirectional edge-states, with the gap Chern number understood as the sum of the Chern numbers of *all* the individual bands below the gap [31].

## B. Continuum case

In the continuum case, the operators $\hat{L}$ and $\mathbf{M}_g$ are independent of all the spatial coordinates. Hence, for plane wave type solutions ($\mathbf{Q} = \mathbf{Q}_{n\mathbf{k}} e^{i\mathbf{k}\cdot\mathbf{r}}$) Eq. (15) reduces to generalized matrix eigenvalue problem, $\hat{L}(\mathbf{k}) \cdot \mathbf{Q}_{n\mathbf{k}} = \omega \mathbf{M}_g \cdot \mathbf{Q}_{n\mathbf{k}}$, with $\mathbf{Q}_{n\mathbf{k}}$ a constant vector. The topological classification is done using $\hat{L}_\mathbf{k} = \hat{L}(\mathbf{k})$ directly in Eq. (8) and taking the set BZ as the entire wave vector space:

$$\mathcal{C} = \frac{-1}{(2\pi)^2} \iint d^2\mathbf{k} \int_{\omega_{gap}-i\infty}^{\omega_{gap}+i\infty} d\omega \, \mathrm{Tr}\left\{ \partial_1 \hat{L}_\mathbf{k} \cdot \mathcal{G}_\mathbf{k} \cdot \partial_2 \hat{L}_\mathbf{k} \cdot \partial_\omega \mathcal{G}_\mathbf{k} \right\}, \tag{19}$$

where $\mathcal{G}_\mathbf{k}(\omega) = i\left(\hat{L}_\mathbf{k} - \mathbf{M}_g \omega\right)^{-1}$ is a matrix.

As already discussed in Sect. II, in order that $\mathcal{C}$ is really topological it is necessary to enforce a high-frequency spatial-cut off in the electromagnetic response [26, 36]. It can be checked that $\mathcal{C}$ can be written in terms of the photonic Green function $\overline{\mathbf{G}}_\mathbf{k}(\omega) = i\left[\hat{N}(\mathbf{k}) - \omega \mathbf{M}(\mathbf{k}, \omega)\right]^{-1}$ as in Eq. (34) of Ref. [26], which thereby remains valid for non-Hermitian photonic systems. Furthermore, for a generic electromagnetic continuum with a high-frequency cut-off the Chern number can be computed using Eq. (41) of Ref. [26], even in presence of material loss. In particular, Eq. (5) of Sect. II (Eq. (43) of Ref. [26]) gives the gap Chern numbers of a non-Hermitian electric gyrotropic material.

## V. Edge states

In regular Hermitian topological systems, the "bulk edge correspondence" establishes a precise link between the Chern numbers of two topological materials and the number of

-22-

edge states supported by a material interface [7, 10, 31, 37, 45, 46, 47]. An illuminating proof of the bulk edge correspondence was recently obtained relying on a link between topological photonics and fluctuation-electrodynamics [31]. Remarkably, it turns out that the thermal (quantum) fluctuation induced light angular momentum spectral density is precisely quantized in a closed topological cavity and that its quantum is precisely the photonic Chern number of the bulk region [31, 48]. The nontrivial angular momentum of thermal light in a closed cavity is due to the circulation of electromagnetic energy in closed orbits. This effect may occur in nonreciprocal photonic systems in thermal equilibrium with a large reservoir [49-51]. The proof of the "bulk edge correspondence" in Ref. [31] relies on the assumption that the material loss is vanishingly small, so that the system dynamics is effectively Hermitian.

Recently, there has been some controversy about the application of the bulk edge correspondence to non-Hermitian systems [10, 15, 16, 17, 19]. Several articles have underlined that the spectrum of a non-Hermitian system with periodic-type boundaries may differ dramatically from the spectrum of the same system with "opaque-type" boundaries (typically referred to as "open" boundaries in the condensed matter literature), i.e., with boundaries that are impenetrable by the wave [10, 15, 16]. Furthermore, some non-Hermitian systems terminated with opaque-type boundaries can have all the states anomalously localized at the boundary, and thereby the closed system states are apparently disconnected from the (Bloch) extended states of the associated periodic system. This property is known as the "non-Hermitian skin effect" [10, 16]. To overcome this problem, a non-Bloch bulk-boundary correspondence was recently developed in Refs. [16, 17, 18], relying on topological invariants defined in a generalized Brillouin zone with a complex-valued wave vector.

In contrast, here we show that –consistent with the *standard* bulk-edge correspondence– the spectrum of non-Hermitian Chern-type topological insulators described by linear differential-equations as in Sect. III.D with $\hat{L}_{\mathbf{k}}$ as in Eq. (16) (e.g., lossy photonic crystals) becomes gapless when the topological material is surrounded by opaque-type walls.

To begin with, we note that similar to our previous article [31] the gap Chern number integral in Eq. (12) can be written in terms of the Green function $\mathcal{G}$ of a cavity that encompasses many unit cells of the photonic crystal as follows (see Appendix D for the details and for the exact definition of $\mathcal{G}$):

$$\mathcal{C} = \lim_{A_{tot} \to \infty} \frac{1}{A_{tot}} \int_{\omega_{\text{gap}}-i\infty}^{\omega_{\text{gap}}+i\infty} d\omega \iint dV dV' \left[ \text{tr} \left\{ \left[ \partial_2 \hat{L} \cdot \mathcal{G}(\mathbf{r}, \mathbf{r}', \omega) \right] \cdot \left[ \partial_1 \hat{L} \cdot \partial_\omega \mathcal{G}(\mathbf{r}', \mathbf{r}, \omega) \right] \right\} \right]. \quad (20)$$

Here, $A_{tot}$ is the area of the cavity cross-section parallel to the *xoy*-plane, which should be large enough so that the discrete spectrum of the cavity approaches the continuum result. The Green function $\mathcal{G}$ satisfies periodic boundary conditions at the cavity walls. The volume integrals are over the entire cavity domain.

The key argument is that the integral in Eq. (20) depends critically on the boundary conditions imposed on the cavity walls [31]. Specifically, in Appendix E it is demonstrated that when the Green function satisfies opaque-type boundary conditions (e.g., for a perfect electric conductor – PEC – boundary) the value of $\mathcal{C}$ calculated with Eq. (20) vanishes. At first sight this result is at odds with the fact that Eq. (20) gives the gap Chern number when the boundary conditions are taken as periodic. Indeed, in a band-gap (i.e., in a vertical strip of the complex frequency plane wherein the bulk region does not support photonic states) the Green function calculated for source ($\mathbf{r}'$) and observation ($\mathbf{r}$) points interior to the cavity must be nearly independent of the boundary conditions imposed on the cavity walls. Note that the Green function $\mathscr{G}(\mathbf{r},\mathbf{r}',\omega)$ must decay exponentially with $|\mathbf{r}-\mathbf{r}'|$ in the bulk region due to the absence of states with a real-valued wave vector.

The only sensible explanation for the critical dependence of the integral (20) on the boundary conditions (periodic vs. opaque) is that the spectrum becomes gapless for opaque-type boundaries, i.e., the opaque boundaries must close the band-gap and lead to the emergence of edge states at the cavity walls [31]. This property suggests that the standard bulk-edge correspondence holds for non-Hermitian systems described by linear differential-equations. A more detailed discussion of the bulk-edge correspondence in non-Hermitian systems is left for future work.

In the following, we illustrate the application of the bulk edge correspondence to an interface ($y=0$) between a gyrotropic material with $\omega_0 = 0.8\omega_p$ (in the semi-space $y>0$) and a perfect electric conductor. The spatial cut-off of the gyrotropic material is taken equal to $k_{max} = 10\omega_p/c$ and the edge states are computed using the formalism of Ref. [37]. Figure 4 represents the edge states dispersion $\omega = \omega(k_x)$ for $k_x$ real-valued and topological materials with collision frequency $\Gamma = 0.5\omega_p$ or $\Gamma = 0$. For simplicity, only the positive-frequency modes (with $\omega' > 0$) are shown. Figure 4a depicts the real part of the edge states natural frequency ($\omega' = \omega'(k_x)$) while Fig. 4b shows the imaginary part of the natural frequency ($\omega'' = \omega''(k_x)$) for $\Gamma = 0.5\omega_p$. For comparison, the dispersion of the bulk states of the gyrotropic material with $\Gamma = 0.5\omega_p$ is also represented in Fig. 4a (solid black lines). Consistent with the bulk-edge correspondence principle, the edge states dispersion (dot-dashed green lines in Fig. 4a) span entirely the two positive frequency band gaps.

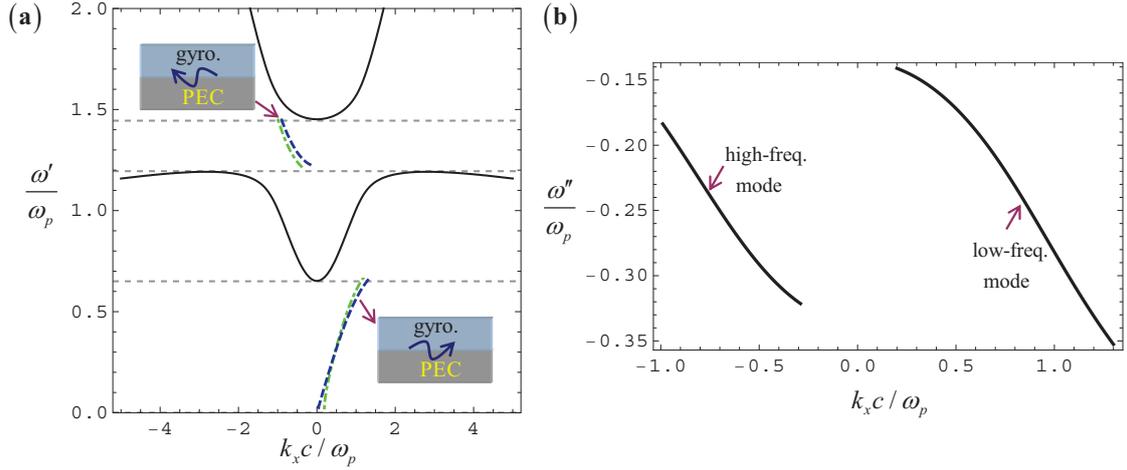

Fig. 4 Dispersion of the complex edge states supported by an interface of a gyrotropic material and a perfect electric conductor. (a) Real part of the edge wave oscillation frequency as a function of $k_x$. The figure only shows the edge-state dispersions in the band-gaps. Dashed blue lines: $\Gamma = 0$ (lossless gyrotropic medium). Dot-dashed green lines: $\Gamma = 0.5\omega_p$ (lossy gyrotropic medium). The solid black lines represent the dispersion ($\omega'$ vs. $k_x$) of the lossy gyrotropic bulk medium; the gray dashed horizontal lines delimit the corresponding band-gaps. (b) Imaginary part of the edge wave oscillation frequencies as a function of $k_x$ for the case $\Gamma = 0.5\omega_p$. The parameters of the gyrotropic material are $\omega_0 = 0.8\omega_p$ and $k_{max} = 10\omega_p / c$.

Figure 5 show the locus of the edge states natural frequencies ($\omega = \omega(k_x)$ with $k_x$ real-valued) in the complex plane (dashed green lines). This alternative representation further highlights that the edge states span the entire gap (i.e., the vertical strips of the complex plane with no bulk modes) and finally merge with the locus of the bulk-material natural frequencies (black curves) in the complex plane. Note that similar to the example of Sect. II, the low frequency gap has topological number −1, while the high-frequency gap has topological number +1. Thus, the bulk edge correspondence predicts correctly the number of edge states in the complex-frequency gaps.

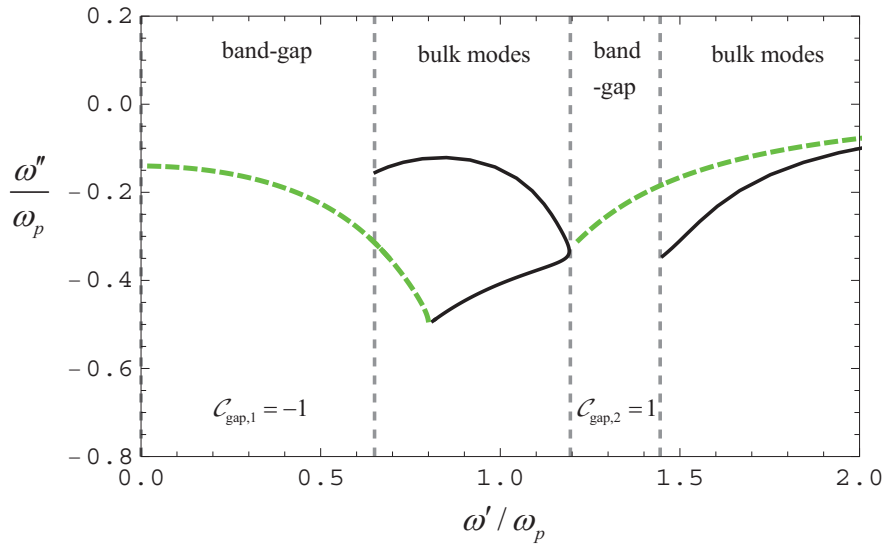

Fig. 5 Locus of the edge states (dashed green lines) and of the bulk modes (black solid lines) of the gyrotropic material in the complex-frequency plane for the example of Fig. 4 with $\Gamma = 0.5\omega_p$. The figure only shows the positive-frequency part of the spectrum.

## VI. Summary

We developed a gauge-independent Green function formalism to calculate the topological invariants of non-Hermitian (fermionic or bosonic) systems, with a focus on photonic platforms. Our analysis shows that the standard Green function methods developed for topological Hermitian systems [26-30] can be extended in a straightforward manner to non-Hermitian platforms, and makes clear that the band gaps in the complex frequency plane must be understood as the regions wherein the system Green function is analytic (e.g., vertical strips of the complex plane that separate the complex eigenfrequencies into two disjoint sets). The Chern number may be found by integrating the Green function along a curve lying in the band-gap that separates the relevant bands in the complex-frequency plane. Furthermore, it was shown that similar to the Hermitian case, the Chern number integral can be expressed in terms of the Green function of a large cavity, and that its value is highly sensitive to the boundary conditions (periodic vs. opaque) imposed on the cavity walls. This property implies that the spectrum of a large topological cavity terminated with opaque-type walls must be gapless. Thus, our analysis suggests that the standard bulk-edge correspondence remains valid in non-Hermitian systems described by linear differential-equations, i.e., that the number of edge-states can be linked to the Chern invariant.

Using the developed theory we characterized the topological phases of electromagnetic continua. In particular, it was shown that magnetized electric plasmas retain their topological properties even in the presence of strong material loss, and that topological edge states with complex-valued frequencies emerge at the interface of the magnetized plasma and a metal wall.

To conclude, it is relevant to note that while in the Hermitian case the photonic Chern number can be understood as the quantum of the fluctuation-induced angular momentum [31, 48], it is not obvious how to generalize such a result to topological platforms with strong material loss. The main obstacle is that different from the Hermitian case [26], for lossy systems the Chern number integral (17) cannot be expressed as an integral over a semi-infinite straight line contained in the upper-half frequency plane. Thereby, it does not seem possible to link the thermal (quantum) fluctuation induced angular momentum of a topological non-Hermitian cavity with the Chern number of the bulk region. The physical meaning of the Chern number of photonic platforms with strong material absorption remains thus an open problem.

**Acknowledgements:** This work is supported in part by the IET under the A F Harvey Engineering Research Prize, by Fundação para a Ciência e a Tecnologia grant number PTDC/EEI-TEL/4543/2014 and by Instituto de Telecomunicações under project UID/EEA/50008/2017.

## Appendix A: Effect of perturbations on the gap Chern number

In this Appendix, we calculate $\partial_\alpha \mathcal{C}$ (Eq. (9) of the main text) with the gap Chern number given by Eq. (8), and $\alpha$ some parameter associated with a deformation of the system Hamiltonian ($\partial_\alpha = \partial/\partial\alpha$). Our analysis extends a result reported in [27, p. 166] to non-Hermitian systems.

To begin with, we note that $\partial_\omega \mathcal{G}_\mathbf{k} = -\mathcal{G}_\mathbf{k} \cdot \partial_\omega \mathcal{G}_\mathbf{k}^{-1} \cdot \mathcal{G}_\mathbf{k}$ and hence Eq. (8) may be rewritten as:

$$\mathcal{C} = \frac{-1}{(2\pi)^2} \iint_{B.Z.} d^2\mathbf{k} \int_{\omega_{\text{gap}}-i\infty}^{\omega_{\text{gap}}+i\infty} d\omega \, \text{Tr}\left\{ \mathcal{G}_\mathbf{k} \cdot \partial_\omega \mathcal{G}_\mathbf{k}^{-1} \cdot \mathcal{G}_\mathbf{k} \cdot \partial_1 \mathcal{G}_\mathbf{k}^{-1} \cdot \mathcal{G}_\mathbf{k} \cdot \partial_2 \mathcal{G}_\mathbf{k}^{-1} \right\}. \tag{A1}$$

By integrating by parts in frequency Eq. (8), it is seen that exchanging the indices "1" and "2" flips the sign of the integral. From this property, it follows the Chern number is given by the symmetrized formula:

$$\mathcal{C} = \frac{-1}{(2\pi)^2} \frac{1}{6} \iint_{B.Z.} d^2\mathbf{k} \int_{\omega_{\text{gap}}-i\infty}^{\omega_{\text{gap}}+i\infty} d\omega \, \varepsilon^{ijl} \text{Tr}\left\{ \mathcal{G} \cdot \partial_i \mathcal{G}^{-1} \cdot \mathcal{G} \cdot \partial_j \mathcal{G}^{-1} \cdot \mathcal{G} \cdot \partial_l \mathcal{G}^{-1} \right\}. \tag{A2}$$

Here, $\varepsilon^{ijl}$ is the Levi-Civita symbol and the summation over $i,j,l \in \{0,1,2\}$ is implicit. Furthermore, by definition $\partial_0 = \partial_\omega$ and to clean up the notations we dropped the index $\mathbf{k}$.

Let us now suppose that the Hamiltonian varies continuously with some generic parameter $\alpha$, so that the operator $\mathcal{G} = \mathcal{G}(\alpha)$, and thereby also the Chern number, can be regarded as functions of $\alpha$. The gap frequency $\omega_{\text{gap}}$ is fixed and should lie in a complete band-gap (vertical strip of the complex plane with no eigenfrequencies), independent of the value of $\alpha$. From $\partial_\alpha \mathcal{G}^{-1} = -\mathcal{G}^{-1} \cdot \partial_\alpha \mathcal{G} \cdot \mathcal{G}^{-1}$ we find that:

$$\partial_\alpha \left( \mathcal{G} \cdot \partial_i \mathcal{G}^{-1} \right) = -\mathcal{G} \cdot \partial_i \mathcal{G}^{-1} \partial_\alpha \mathcal{G} \cdot \mathcal{G}^{-1} - \partial_\alpha \partial_i \mathcal{G} \cdot \mathcal{G}^{-1}. \tag{A3}$$

Using the cyclic property of the trace, it is straightforward to show that the derivative with respect to $\alpha$ of a generic term of the integrand of Eq. (A2) is:

$$\partial_\alpha \text{Tr}\left\{ \left( \mathcal{G} \cdot \partial_i \mathcal{G}^{-1} \right) \cdot \left( \mathcal{G} \cdot \partial_j \mathcal{G}^{-1} \right) \cdot \left( \mathcal{G} \cdot \partial_l \mathcal{G}^{-1} \right) \right\}$$
$$= - \sum_{\substack{(\mu,\nu,o)=(i,j,l),\\(j,l,i),(l,i,j)}} \text{Tr}\left\{ \left( \partial_\alpha \mathcal{G} \cdot \mathcal{G}^{-1} \right) \cdot \left( \mathcal{G} \cdot \partial_\mu \mathcal{G}^{-1} \right) \cdot \left( \mathcal{G} \cdot \partial_\nu \mathcal{G}^{-1} \right) \cdot \left( \mathcal{G} \cdot \partial_o \mathcal{G}^{-1} \right) \right\} \tag{A4}$$
$$- \sum_{\substack{(\mu,\nu,o)=(i,j,l),\\(j,l,i),(l,i,j)}} \text{Tr}\left\{ \partial_\mu \partial_\alpha \mathcal{G} \cdot \mathcal{G}^{-1} \cdot \left( \mathcal{G} \cdot \partial_\nu \mathcal{G}^{-1} \right) \cdot \left( \mathcal{G} \cdot \partial_o \mathcal{G}^{-1} \right) \right\}$$

The sums are over 3 terms: $(i,j,l)$, $(j,l,i)$ and $(l,i,j)$. From here it follows that:

$$\partial_\alpha \text{Tr}\left\{ \left( \mathcal{G} \cdot \partial_i \mathcal{G}^{-1} \right) \cdot \left( \mathcal{G} \cdot \partial_j \mathcal{G}^{-1} \right) \cdot \left( \mathcal{G} \cdot \partial_l \mathcal{G}^{-1} \right) \right\} =$$
$$- \sum_{\substack{(\mu,\nu,o)=(i,j,l),\\(j,l,i),(l,i,j)}} \partial_\mu \text{Tr}\left\{ \left( \partial_\alpha \mathcal{G} \cdot \mathcal{G}^{-1} \right) \cdot \left( \mathcal{G} \cdot \partial_\nu \mathcal{G}^{-1} \right) \cdot \left( \mathcal{G} \cdot \partial_o \mathcal{G}^{-1} \right) \right\} \tag{A5}$$
$$+ \sum_{\substack{(\mu,\nu,o)=(i,j,l),\\(j,l,i),(l,i,j)}} \text{Tr}\left\{ \left( \partial_\alpha \mathcal{G} \cdot \mathcal{G}^{-1} \right) \cdot \partial_\mu \left[ \left( \mathcal{G} \cdot \partial_\nu \mathcal{G}^{-1} \right) \cdot \left( \mathcal{G} \cdot \partial_o \mathcal{G}^{-1} \right) \right] \right\}$$

Next, we differentiate both members of Eq. (A2) with respect to $\alpha$ and use the above result. Noting that $\sum_{i,j,l} \varepsilon^{ijl} \sum_{\substack{(\mu,\nu,o)=(i,j,l),\\(j,l,i),(l,i,j)}} a_{\mu\nu o} = 3 \sum_{i,j,l} \varepsilon^{ijl} a_{ijl}$ for arbitrary coefficients $a_{ijl}$, we are left with (the summation over all $i,j,l = 0,1,2$ is implicit):

$$\partial_\alpha C = \frac{-1}{(2\pi)^2} \frac{1}{2} \iint_{B.Z.} d^2\mathbf{k} \int_{\omega_{gap}-i\infty}^{\omega_{gap}+i\infty} d\omega\, \varepsilon^{ijl} \left[ -\partial_i \text{Tr}\left\{ \partial_\alpha \mathcal{G} \cdot \partial_j \mathcal{G}^{-1} \cdot \mathcal{G} \cdot \partial_l \mathcal{G}^{-1} \right\} + \right.$$
$$\left. \text{Tr}\left\{ \left( \partial_\alpha \mathcal{G} \cdot \mathcal{G}^{-1} \right) \cdot \partial_i \left[ \left( \mathcal{G} \cdot \partial_j \mathcal{G}^{-1} \right) \cdot \left( \mathcal{G} \cdot \partial_l \mathcal{G}^{-1} \right) \right] \right\} \right] \qquad (A6)$$

Now, we observe that

$$\partial_i \left[ \left( \mathcal{G} \cdot \partial_j \mathcal{G}^{-1} \right) \cdot \left( \mathcal{G} \cdot \partial_l \mathcal{G}^{-1} \right) \right]$$
$$= \left( \partial_i \mathcal{G} \cdot \partial_j \mathcal{G}^{-1} \cdot \mathcal{G} + \mathcal{G} \cdot \partial_j \mathcal{G}^{-1} \partial_i \mathcal{G} \right) \cdot \partial_l \mathcal{G}^{-1} + \left( \mathcal{G} \cdot \partial_i \partial_j \mathcal{G}^{-1} \cdot \mathcal{G} \cdot \partial_l \mathcal{G}^{-1} \right) + \left( \mathcal{G} \cdot \partial_j \mathcal{G}^{-1} \cdot \mathcal{G} \cdot \partial_i \partial_l \mathcal{G}^{-1} \right)$$
(A7)

Using $\mathcal{G} \cdot \partial_j \mathcal{G}^{-1} \cdot \partial_i \mathcal{G} \cdot \mathcal{G}^{-1} = \partial_j \mathcal{G} \cdot \partial_i \mathcal{G}^{-1}$ we may further write:

$$\partial_i \left[ \left( \mathcal{G} \cdot \partial_j \mathcal{G}^{-1} \right) \cdot \left( \mathcal{G} \cdot \partial_l \mathcal{G}^{-1} \right) \right]$$
$$= \left( \partial_i \mathcal{G} \cdot \partial_j \mathcal{G}^{-1} + \partial_j \mathcal{G} \cdot \partial_i \mathcal{G}^{-1} \right) \cdot \mathcal{G} \cdot \partial_l \mathcal{G}^{-1} + \left( \mathcal{G} \cdot \partial_i \partial_j \mathcal{G}^{-1} \cdot \mathcal{G} \cdot \partial_l \mathcal{G}^{-1} \right) + \left( \mathcal{G} \cdot \partial_j \mathcal{G}^{-1} \cdot \mathcal{G} \cdot \partial_i \partial_l \mathcal{G}^{-1} \right)$$
(A8)

The first and second terms on the right-hand side are invariant under a permutation of the indices $i, j$, whereas the third term is invariant under a permutation of the indices $i, l$. These properties imply that $\sum_{i,j,l} \varepsilon^{ijl} \partial_i \left[ \left( \mathcal{G} \cdot \partial_j \mathcal{G}^{-1} \right) \cdot \left( \mathcal{G} \cdot \partial_l \mathcal{G}^{-1} \right) \right] = 0$, and therefore the second term in the integrand of Eq. (A6) vanishes. This observation yields (restoring the $\mathbf{k}$ index) Eq. (9) of the main text, which is the desired result.

## Appendix B: The gap Chern number integral

In the following, we derive Eq. (10) of the main text relying on the matrix representation $\hat{\mathcal{L}}_\mathbf{k} \to S_\mathbf{k} \cdot \Omega_\mathbf{k} \cdot S_\mathbf{k}^{-1}$ of the operator $\hat{\mathcal{L}}_\mathbf{k}$.

To begin with, we note that from $\hat{\mathcal{L}}_\mathbf{k} = \mathbf{M}_g^{-1/2} \hat{L}_\mathbf{k} \mathbf{M}_g^{-1/2}$ the Green function operator can be written as $\mathcal{G}_\mathbf{k} = \mathbf{M}_g^{-1/2} \tilde{\mathcal{G}}_\mathbf{k} \mathbf{M}_g^{-1/2}$ with $\tilde{\mathcal{G}}_\mathbf{k} = i\left( \hat{L}_\mathbf{k} - \omega \mathbf{1} \right)^{-1}$. Taking into account that $\mathbf{M}_g$ is independent of the wave vector and using the cyclic property of the trace operator ($\mathrm{Tr}\{\mathbf{A}\cdot\mathbf{B}\} = \mathrm{Tr}\{\mathbf{B}\cdot\mathbf{A}\}$) it can be easily checked that Eq. (8) still holds with $\tilde{\mathcal{G}}_{\mathbf{k}}$ in the place of $\mathcal{G}_{\mathbf{k}}$. Clearly, $\tilde{\mathcal{G}}_{\mathbf{k}} = i\left(\hat{\mathcal{L}}_{\mathbf{k}} - \omega\mathbf{1}\right)^{-1}$ is represented by the matrix $\tilde{\mathcal{G}}_{\mathbf{k}} \to i S_{\mathbf{k}} \cdot (\Omega_{\mathbf{k}} - \omega\mathbf{1})^{-1} \cdot S_{\mathbf{k}}^{-1}$. Substituting this result into Eq. (8) (with $\tilde{\mathcal{G}}_{\mathbf{k}}$ in the place of $\mathcal{G}_{\mathbf{k}}$) and noting that $(\Omega_{\mathbf{k}} - \omega\mathbf{1})^{-1}$ is a diagonal matrix it is found that:

$$\mathcal{C} = \frac{-1}{(2\pi)^2} \iint_{B.Z.} d^2\mathbf{k} \int_{\omega_{\mathrm{gap}} - i\infty}^{\omega_{\mathrm{gap}} + i\infty} d\omega\, \mathrm{Tr}\left\{ S_{\mathbf{k}}^{-1} \cdot \partial_1 \tilde{\mathcal{G}}_{\mathbf{k}}^{-1} \cdot S_{\mathbf{k}} \cdot (\Omega_{\mathbf{k}} - \mathbf{1}\omega)^{-1} \cdot S_{\mathbf{k}}^{-1} \cdot \partial_2 \tilde{\mathcal{G}}_{\mathbf{k}}^{-1} \cdot S_{\mathbf{k}} \cdot (\Omega_{\mathbf{k}} - \mathbf{1}\omega)^{-2} \right\}$$

(B1)

Straightforward calculations show that $i S_{\mathbf{k}}^{-1} \cdot \partial_j \tilde{\mathcal{G}}_{\mathbf{k}}^{-1} \cdot S_{\mathbf{k}} = \left[ S_{\mathbf{k}}^{-1} \partial_j S_{\mathbf{k}}, \Omega_{\mathbf{k}} \right] + \partial_j \Omega_{\mathbf{k}}$ with $[\mathbf{A}, \mathbf{B}] = \mathbf{A} \cdot \mathbf{B} - \mathbf{B} \cdot \mathbf{A}$ the commutator of two operators. Hence, the gap Chern number may be written as:

$$\mathcal{C} = \frac{1}{(2\pi)^2} \iint_{B.Z.} d^2\mathbf{k} \int_{\omega_{\mathrm{gap}} - i\infty}^{\omega_{\mathrm{gap}} + i\infty} d\omega$$
$$\mathrm{Tr}\left\{ \left( \left[ S_{\mathbf{k}}^{-1} \cdot \partial_1 S_{\mathbf{k}}, \Omega_{\mathbf{k}} \right] + \partial_1 \Omega_{\mathbf{k}} \right) \cdot (\Omega_{\mathbf{k}} - \mathbf{1}\omega)^{-1} \cdot \left( \left[ S_{\mathbf{k}}^{-1} \cdot \partial_2 S_{\mathbf{k}}, \Omega_{\mathbf{k}} \right] + \partial_2 \Omega_{\mathbf{k}} \right) \cdot (\Omega_{\mathbf{k}} - \mathbf{1}\omega)^{-2} \right\}$$

(B2)

The operator inside the trace can be written as a sum of four terms. The 3 terms that depend explicitly on $\partial_j \Omega_{\mathbf{k}}$ vanish. For example, the term $\mathrm{Tr}\left\{ \partial_1 \Omega_{\mathbf{k}} \cdot (\Omega_{\mathbf{k}} - \mathbf{1}\omega)^{-1} \cdot \left[ S_{\mathbf{k}}^{-1} \cdot \partial_2 S_{\mathbf{k}}, \Omega_{\mathbf{k}} \right] \cdot (\Omega_{\mathbf{k}} - \mathbf{1}\omega)^{-2} \right\}$ can be rewritten as (using the cyclic property of the trace and noting that the matrices $\partial_j \Omega_{\mathbf{k}}$ and $(\Omega_{\mathbf{k}} - \mathbf{1}\omega)^{-m}$ are diagonal and hence commute) $\mathrm{Tr}\left\{ \partial_1 \Omega_{\mathbf{k}} \cdot \left[ S_{\mathbf{k}}^{-1} \cdot \partial_2 S_{\mathbf{k}}, \Omega_{\mathbf{k}} \right] \cdot (\Omega_{\mathbf{k}} - \mathbf{1}\omega)^{-3} \right\}$. The only factor that depends on frequency is $(\Omega_{\mathbf{k}} - \mathbf{1}\omega)^{-3}$. The integral $\int_{\omega_{gap}-i\infty}^{\omega_{gap}+i\infty} d\omega (\Omega_{\mathbf{k}} - \mathbf{1}\omega)^{-3}$ vanishes because the residues of all poles vanish. Thus, the considered term does not contribute to the Chern number. Hence, we are left with:

$$\mathcal{C} = \frac{1}{(2\pi)^2} \iint_{B.Z.} d^2\mathbf{k} \int_{\omega_{gap}-i\infty}^{\omega_{gap}+i\infty} d\omega \, \mathrm{Tr}\left\{\left[S_{\mathbf{k}}^{-1} \cdot \partial_1 S_{\mathbf{k}}, \Omega_{\mathbf{k}}\right] \cdot (\Omega_{\mathbf{k}} - \mathbf{1}\omega)^{-1} \cdot \left[S_{\mathbf{k}}^{-1} \cdot \partial_2 S_{\mathbf{k}}, \Omega_{\mathbf{k}}\right] \cdot (\Omega_{\mathbf{k}} - \mathbf{1}\omega)^{-2}\right\}$$

(B3)

To proceed further we use the auxiliary result [26]

$$\int_{\omega_{gap}-i\infty}^{\omega_{gap}+i\infty} d\omega \frac{1}{(\omega - \omega_m)^2} \frac{1}{\omega - \omega_n} = \frac{2\pi i}{(\omega_m - \omega_n)^2} \mathrm{sgn}\left(\omega_{gap} - \mathrm{Re}\{\omega_n\}\right),$$

(B4)

which holds when $\omega_m, \omega_n$ are on different semi-planes relative to the vertical line $\mathrm{Re}\{\omega\} = \omega_{gap}$ (e.g., $\mathrm{Re}\{\omega_m\} > \omega_{gap}$ and $\mathrm{Re}\{\omega_n\} < \omega_{gap}$). The same integral vanishes identically when $\omega_m, \omega_n$ lie on the same semi-plane. Writing $\Omega_{\mathbf{k}} = \sum_n \omega_{n\mathbf{k}} \hat{\mathbf{u}}_n \otimes \hat{\mathbf{u}}_n$ with $\hat{\mathbf{u}}_n$ a vector whose $n$-th element is "1" and with all the other elements identically to zero, it is found that:

$$\mathcal{C} = \frac{1}{(2\pi)^2} \iint_{B.Z.} d^2\mathbf{k} \int_{\omega_{gap}-i\infty}^{\omega_{gap}+i\infty} d\omega \, \mathrm{Tr}\left\{\left[S_{\mathbf{k}}^{-1} \cdot \partial_1 S_{\mathbf{k}}, \Omega_{\mathbf{k}}\right] \cdot \sum_n \frac{-1}{\omega - \omega_{n\mathbf{k}}} \hat{\mathbf{u}}_n \otimes \hat{\mathbf{u}}_n \cdot \left[S_{\mathbf{k}}^{-1} \cdot \partial_2 S_{\mathbf{k}}, \Omega_{\mathbf{k}}\right] \cdot \sum_m \frac{1}{(\omega - \omega_{m\mathbf{k}})^2} \hat{\mathbf{u}}_m \otimes \hat{\mathbf{u}}_m\right\}$$

$$= \frac{1}{2\pi} \iint_{B.Z.} d^2\mathbf{k} \, \mathrm{Tr}\left\{i \sum_{n \in F, m \in E} \frac{-1}{(\omega_{n\mathbf{k}} - \omega_{m\mathbf{k}})^2} \left[S_{\mathbf{k}}^{-1} \cdot \partial_1 S_{\mathbf{k}}, \Omega_{\mathbf{k}}\right] \cdot \hat{\mathbf{u}}_n \otimes \hat{\mathbf{u}}_n \cdot \left[S_{\mathbf{k}}^{-1} \cdot \partial_2 S_{\mathbf{k}}, \Omega_{\mathbf{k}}\right] \cdot \hat{\mathbf{u}}_m \otimes \hat{\mathbf{u}}_m\right\}$$

$$+ \mathrm{Tr}\left\{i \sum_{n \in E, m \in F} \frac{+1}{(\omega_{n\mathbf{k}} - \omega_{m\mathbf{k}})^2} \left[S_{\mathbf{k}}^{-1} \cdot \partial_1 S_{\mathbf{k}}, \Omega_{\mathbf{k}}\right] \cdot \hat{\mathbf{u}}_n \otimes \hat{\mathbf{u}}_n \cdot \left[S_{\mathbf{k}}^{-1} \cdot \partial_2 S_{\mathbf{k}}, \Omega_{\mathbf{k}}\right] \cdot \hat{\mathbf{u}}_m \otimes \hat{\mathbf{u}}_m\right\}$$

(B5)

In the above $E$ and $F$ represent the sets of "empty" and "filled" bands, respectively. After some simplifications we may simply write:

$$\mathcal{C} = \frac{1}{2\pi} \iint_{B.Z.} d^2\mathbf{k} \, i \text{Tr}\left\{ S_\mathbf{k}^{-1} \cdot \partial_1 S_\mathbf{k} \cdot \mathbf{1}_F \cdot S_\mathbf{k}^{-1} \cdot \partial_2 S_\mathbf{k} \cdot \mathbf{1}_E - S_\mathbf{k}^{-1} \cdot \partial_1 S_\mathbf{k} \cdot \mathbf{1}_E \cdot S_\mathbf{k}^{-1} \cdot \partial_2 S_\mathbf{k} \cdot \mathbf{1}_F \right\}, \qquad (B6)$$

where $\mathbf{1}_F = \sum_{n \in F} \hat{\mathbf{u}}_n \otimes \hat{\mathbf{u}}_n$ and $\mathbf{1}_E = \sum_{n \in E} \hat{\mathbf{u}}_n \otimes \hat{\mathbf{u}}_n = \mathbf{1} - \mathbf{1}_F$ are diagonal matrices. Specifically, since the eigenvectors are ordered so that the filled bands are associated with the indices $n = 1, 2, \ldots N_F$, the matrix $\mathbf{1}_F$ has diagonal elements identical to "1" for $n = 1, 2, \ldots N_F$ and identical to "0" otherwise. Note that $\mathbf{1}_E$ and $\mathbf{1}_F$ are independent of the wave vector. The term associated with the trace can be written in a more compact manner as $\text{Tr}\{\ldots\} = \text{Tr}\{S_\mathbf{k}^{-1} \cdot \partial_1 S_\mathbf{k} \cdot \mathbf{1}_F \cdot S_\mathbf{k}^{-1} \cdot \partial_2 S_\mathbf{k} \cdot \mathbf{1}_E - 1 \leftrightarrow 2\}$ where $1 \leftrightarrow 2$ represents the first term with the indices "1" and "2" exchanged. Noting that $\text{Tr}\{S_\mathbf{k}^{-1} \cdot \partial_1 S_\mathbf{k} \cdot \mathbf{1}_F \cdot S_\mathbf{k}^{-1} \cdot \partial_2 S_\mathbf{k} \cdot \mathbf{1}_F - 1 \leftrightarrow 2\} = 0$ one finds that:

$$\mathcal{C} = \frac{1}{2\pi} \iint_{B.Z.} d^2\mathbf{k} \, i \text{Tr}\left\{ S_\mathbf{k}^{-1} \cdot \partial_1 S_\mathbf{k} \cdot \mathbf{1}_F \cdot S_\mathbf{k}^{-1} \cdot \partial_2 S_\mathbf{k} - 1 \leftrightarrow 2 \right\}. \qquad (B7)$$

Using the cyclic properties of the trace and $\partial_2 S_\mathbf{k}^{-1} = -S_\mathbf{k}^{-1} \cdot \partial_2 S_\mathbf{k} \cdot S_\mathbf{k}^{-1}$, one gets Eq. (10) of the main text. Notice that $\mathcal{C}$ is fully independent of the eigenvalues ($\omega_{n\mathbf{k}}$) of the operator: it only depends on the matrix ($S_\mathbf{k}$) that represents the eigenvectors on a fixed basis.

In the rest of this Appendix, we discuss how a gauge transformation affects the Berry potential $\mathcal{A}_\mathbf{k} = i \text{Tr}\{S_\mathbf{k}^{-1} \cdot \partial_\mathbf{k} S_\mathbf{k} \cdot \mathbf{1}_F\}$. Specifically, consider two generic bases of eigenfunctions $\mathcal{Q}_{n\mathbf{k}}$ and $\mathcal{Q}'_{n\mathbf{k}}$, with the elements $n = 1, 2, \ldots, N_F$ generating the eigenspace of filled bands, and the remaining elements generating the eigenspace of empty bands. Then, the coordinates of $\mathcal{Q}'_{n\mathbf{k}}$ in the $\mathcal{Q}_{n\mathbf{k}}$ basis are determined by a matrix of the form $\mathbf{s}_{F\mathbf{k}} + \mathbf{s}_{E\mathbf{k}}$ with $\mathbf{s}_{F\mathbf{k}} = \mathbf{1}_F \cdot \mathbf{s}_{F\mathbf{k}} \cdot \mathbf{1}_F$ and $\mathbf{s}_{E\mathbf{k}} = \mathbf{1}_E \cdot \mathbf{s}_{E\mathbf{k}} \cdot \mathbf{1}_E$ (note that the only nontrivial elements of the matrix $[\mathbf{s}_{F\mathbf{k}}]_{m,n}$ are those with $m, n = 1, 2, ..., N_F$; this property implies that $(\mathbf{s}_{F\mathbf{k}} + \mathbf{s}_{E\mathbf{k}})^{-1} = \mathbf{s}_{F\mathbf{k}}^{-1} + \mathbf{s}_{E\mathbf{k}}^{-1}$; for non-degenerate eigenvalues the matrix $\mathbf{s}_{F\mathbf{k}} + \mathbf{s}_{E\mathbf{k}}$ is typically diagonal). From here, it follows that the matrix $S'_\mathbf{k}$ with the coordinates of $\mathcal{Q}'_{n\mathbf{k}}$ in the basis $\mathbf{e}_1, \mathbf{e}_2, ...$ is $S'_\mathbf{k} = S_\mathbf{k} (\mathbf{s}_{E\mathbf{k}} + \mathbf{s}_{F\mathbf{k}})$. Therefore, the Berry potential is transformed as:

$$\mathcal{A}'_\mathbf{k} = i \operatorname{Tr}\left\{ (\mathbf{s}_{E\mathbf{k}}^{-1} + \mathbf{s}_{F\mathbf{k}}^{-1}) \cdot S_\mathbf{k}^{-1} \cdot \partial_\mathbf{k} [S_\mathbf{k} (\mathbf{s}_{E\mathbf{k}} + \mathbf{s}_{F\mathbf{k}})] \cdot \mathbf{1}_F \right\}. \tag{B8}$$

Noting that $\mathbf{s}_{E\mathbf{k}} \cdot \mathbf{1}_F = 0 = \mathbf{s}_{E\mathbf{k}}^{-1} \cdot \mathbf{1}_F$ it is readily found that:

$$\mathcal{A}'_\mathbf{k} = \mathcal{A}_\mathbf{k} + i \operatorname{Tr}\left\{ \mathbf{s}_{F\mathbf{k}}^{-1} \cdot \partial_\mathbf{k} \mathbf{s}_{F\mathbf{k}} \right\}. \tag{B9}$$

Using the general property $\operatorname{Tr}\left\{ \mathbf{s}_{F\mathbf{k}}^{-1} \cdot \partial_\mathbf{k} \mathbf{s}_{F\mathbf{k}} \right\} = \partial_\mathbf{k} \ln \det \mathbf{s}_{F\mathbf{k}}$ (this result can be readily proven when $\mathbf{s}_{F\mathbf{k}}$ is a diagonal matrix, i.e., in case of non-degenerate eigenfunctions; the formula holds true even if $\mathbf{s}_{F\mathbf{k}}$ is not diagonal) it is found that:

$$\mathcal{A}'_\mathbf{k} = \mathcal{A}_\mathbf{k} + \partial_\mathbf{k} (i \ln \det \mathbf{s}_{F\mathbf{k}}). \tag{B10}$$

This yields Eq. (11) of the main text with $g_\mathbf{k} = \det \mathbf{s}_{F\mathbf{k}}$.

## Appendix C: Electrodynamics of non-Hermitian dispersive systems

In this Appendix, it is shown that the electrodynamics of generic non-Hermitian dispersive systems can be reformulated as a Schrödinger-type time-evolution problem.

We consider generic bianisotropic materials with a frequency domain response determined by a 6×6 material matrix $\mathbf{M}$ that links the electromagnetic field vectors $\mathbf{f} = (\mathbf{E} \quad \mathbf{H})^T$ and $\mathbf{g} = (\mathbf{D} \quad \mathbf{B})^T$ as follows:

$$\mathbf{g}(\mathbf{r},\omega) = \mathbf{M}(\mathbf{r},\omega) \cdot \mathbf{f}(\mathbf{r},\omega). \tag{C1}$$

It is assumed that $\mathbf{M}$ is a meromorphic function in the complex plane, so that it has a partial-fraction decomposition of the form:

$$\mathbf{M}(\mathbf{r},\omega) = \mathbf{M}_\infty + \sum_\alpha \frac{(\text{Res}\,\mathbf{M})_\alpha}{\omega - \omega_{p,\alpha}}. \tag{C2}$$

Here, $\mathbf{M}_\infty = \lim_{\omega \to \infty} \mathbf{M}(\omega)$ gives the asymptotic high-frequency response of the material, $\omega_{p,\alpha}$ are the (complex-valued) poles of $\mathbf{M}$, all in the lower-half plane for passive systems, and $(\text{Res}\,\mathbf{M})_\alpha = \lim_{\omega \to \omega_{p,\alpha}} \mathbf{M}(\mathbf{r},\omega)(\omega - \omega_{p,\alpha})$ gives the residue of the $\omega_{p,\alpha}$ pole. The reality of the electromagnetic fields imposes the additional constraint $\mathbf{M}(\omega) = \mathbf{M}^*(-\omega^*)$. Let us introduce the auxiliary variables

$$\mathbf{Q}^{(\alpha)}(\mathbf{r},\omega) = \frac{(s_\alpha \omega_{p,\alpha})^{1/2}}{(\omega - \omega_{p,\alpha})} \mathbf{A}_\alpha \cdot \mathbf{f}(\mathbf{r},\omega), \tag{C3}$$

with $s_\alpha = \text{sgn}(\text{Re}\{\omega_{p,\alpha}\})$ and $\mathbf{A}_\alpha = [-s_\alpha (\text{Res}\,\mathbf{M})_\alpha]^{1/2}$. Then, similar to the lossless case [26, 36, 39], it is possible to show that the time-evolution of the state-vector $\mathbf{Q} = \begin{pmatrix} \mathbf{f} & \mathbf{Q}^{(1)} & \ldots & \mathbf{Q}^{(\alpha)} & \ldots \end{pmatrix}^T$ is determined by the differential equation,

$$\hat{L} \cdot \mathbf{Q}(\mathbf{r},t) = i\frac{\partial}{\partial t} \mathbf{M}_g \cdot \mathbf{Q}(\mathbf{r},t) + i\mathbf{j}_g(\mathbf{r},t), \tag{C4}$$

with

$$\hat{L} = \begin{pmatrix} \hat{N} + \sum_\alpha s_\alpha \mathbf{A}_\alpha^2 & (s_1 \omega_{p,1})^{1/2} \mathbf{A}_1 & (s_2 \omega_{p,2})^{1/2} \mathbf{A}_2 & \ldots \\ (s_1 \omega_{p,1})^{1/2} \mathbf{A}_1 & \omega_{p,1}\mathbf{1} & 0 & \ldots \\ (s_2 \omega_{p,2})^{1/2} \mathbf{A}_2 & 0 & \omega_{p,2}\mathbf{1} & \ldots \\ \ldots & \ldots & \ldots & \ldots \end{pmatrix}, \quad \mathbf{M}_g = \begin{pmatrix} \mathbf{M}_\infty & 0 & 0 & \ldots \\ 0 & \mathbf{1} & 0 & \ldots \\ 0 & 0 & \mathbf{1} & \ldots \\ \ldots & \ldots & \ldots & \ldots \end{pmatrix}. \tag{C5}$$

In the above, $\mathbf{j}_g = \begin{pmatrix} \mathbf{j} & 0 & 0 & ... \end{pmatrix}^T$ is a generalized current written in terms of the electric and magnetic current densities, $\mathbf{j} = \begin{pmatrix} \mathbf{j}_e & \mathbf{j}_m \end{pmatrix}^T$. The differential operator $\hat{N}$ is defined as in the main text [Eq. (14)]. For simplicity, the same symbols are used to denote the frequency domain and the time domain fields.

## Appendix D: Chern number as a function of the Green function of a large cavity

In this Appendix, we prove that analogous to the case of Hermitian systems [26], the Chern number can be computed from the Green function of a large cavity satisfying periodic boundary conditions. Specifically, we will show that the gap Chern number in Eq. (12) with $\hat{L}_\mathbf{k}$ given by Eq. (16) can be written as:

$$\mathcal{C} = \lim_{A_{tot} \to \infty} \frac{1}{A_{tot}} \int_{\omega_{gap}-i\infty}^{\omega_{gap}+i\infty} d\omega \iint dV dV' \left[ \text{tr} \left\{ \left[ \partial_2 \hat{L} \cdot \mathcal{G}(\mathbf{r},\mathbf{r}',\omega) \right] \cdot \left[ \partial_1 \hat{L} \cdot \partial_\omega \mathcal{G}(\mathbf{r}',\mathbf{r},\omega) \right] \right\} \right]. \quad \text{(D1)}$$

Here, $\mathcal{G}$ stands for the Green function that satisfies

$$\hat{L}(\mathbf{r},-i\nabla) \cdot \mathcal{G}(\mathbf{r},\mathbf{r}',\omega) = \omega \mathbf{M}_g(\mathbf{r}) \cdot \mathcal{G}(\mathbf{r},\mathbf{r}',\omega) + i\mathbf{1}\delta(\mathbf{r}-\mathbf{r}') \quad \text{(D2)}$$

in a domain that encompasses many (let us say $N_x \times N_y$) unit cells of the photonic crystal (referred from here on as the "cavity") and $A_{tot}$ is the area of the domain cross-section parallel to the *xoy*-plane. The function $\mathcal{G}$ satisfies periodic boundary conditions on the boundaries of the "cavity". Similar to Eq. (12), the operator $\partial_1 \hat{L}$ ($\partial_2 \hat{L}$) acts on the primed (unprimed) coordinates, and by definition

$$\partial_j \hat{L} \equiv \frac{1}{i}\left[ x_j, \hat{L} \right] = \frac{1}{i}\left( x_j \hat{L}(-i\nabla) - \hat{L}(-i\nabla) x_j \right), \quad j=1,2. \quad \text{(D3)}$$

It is also possible to write $\partial_i \hat{L} = \partial_i \hat{L}_{\mathbf{k}=0}$ with $\hat{L}_{\mathbf{k}}$ defined as in Eq. (16).

To begin with, let us introduce $\delta_p(\mathbf{r}) = \sum_{\mathbf{r}_\mathbf{I}} \delta(\mathbf{r} - \mathbf{r}_\mathbf{I})$ with $\mathbf{r}_\mathbf{I} = i_1 a_1 \hat{\mathbf{x}} + i_2 a_2 \hat{\mathbf{y}}$ a generic lattice point inside the relevant cavity ($i_1 = 0,...,N_x - 1, i_2 = 0,...,N_y - 1$), and $\delta$ understood as the $\delta$-function for the large cavity. Clearly, $\delta_p(\mathbf{r})$ is a periodic function and its irreducible domain is coincident with the unit cell of the photonic crystal. It is straightforward to verify that $\frac{1}{N_x N_y} \sum_{\mathbf{k}_\mathbf{J}} \delta_p(\mathbf{r}) e^{i \mathbf{k}_\mathbf{J} \cdot \mathbf{r}} = \delta(\mathbf{r})$ where $\mathbf{k}_\mathbf{J} = j_1 \frac{2\pi}{N_x a_1} \hat{\mathbf{x}} + j_2 \frac{2\pi}{N_x a_2} \hat{\mathbf{y}}$ ($j_1 = 0,...,N_x - 1$, $j_2 = 0,...,N_y - 1$). Using this identity we see that $\mathcal{G}(\mathbf{r},\mathbf{r}') = \frac{1}{N_x N_y} \sum_{\mathbf{k}_\mathbf{J}} \tilde{\mathcal{G}}_{\mathbf{k}_\mathbf{J}}(\mathbf{r},\mathbf{r}')$ where $\tilde{\mathcal{G}}_{\mathbf{k}_\mathbf{J}}$ satisfies the same differential equation as $\mathcal{G}$ but with $\delta_p(\mathbf{r} - \mathbf{r}') e^{i\mathbf{k}_\mathbf{J} \cdot (\mathbf{r} - \mathbf{r}')}$ replacing $\delta(\mathbf{r} - \mathbf{r}')$. By comparison with Eq. (13), it follows that $\tilde{\mathcal{G}}_{\mathbf{k}_\mathbf{J}} = \mathcal{G}_{\mathbf{k}_\mathbf{J}} e^{i\mathbf{k}_\mathbf{J} \cdot (\mathbf{r} - \mathbf{r}')}$. Hence, Eq. (D1) is equivalent to (note that $e^{-i\mathbf{k} \cdot \mathbf{r}'} \partial_1 \hat{L} e^{i\mathbf{k} \cdot \mathbf{r}'} = \partial_1 \hat{L}_\mathbf{k}$ and $e^{-i\mathbf{k} \cdot \mathbf{r}} \partial_2 \hat{L} e^{i\mathbf{k} \cdot \mathbf{r}} = \partial_2 \hat{L}_\mathbf{k}$):

$$\mathcal{C} = \lim_{A_{tot} \to \infty} \frac{1}{A_{tot}} \frac{1}{(N_x N_y)^2} \int_{\omega_{gap} - i\infty}^{\omega_{gap} + i\infty} d\omega \times \\ \iint dV dV' \sum_{\mathbf{k}_\mathbf{J}, \mathbf{k}_{\mathbf{J}'}} e^{i(\mathbf{k}_\mathbf{J} - \mathbf{k}_{\mathbf{J}'}) \cdot (\mathbf{r} - \mathbf{r}')} \left[ \text{tr} \left\{ \left[ \partial_2 \hat{L}_\mathbf{k} \cdot \mathcal{G}_{\mathbf{k}_\mathbf{J}}(\mathbf{r},\mathbf{r}',\omega) \right] \cdot \left[ \partial_1 \hat{L}_\mathbf{k} \cdot \partial_\omega \mathcal{G}_{\mathbf{k}_{\mathbf{J}'}}(\mathbf{r}',\mathbf{r},\omega) \right] \right\} \right]$$
(D4)

The term in rectangular brackets is periodic over the (space domain) unit cell. Thus, after integration over the entire cavity only the terms with $\mathbf{k}_\mathbf{J} = \mathbf{k}_{\mathbf{J}'}$ survive. Therefore, after reducing the integration to a single unit cell we are left with:

$$\mathcal{C} = \lim_{A_{tot} \to \infty} \frac{1}{A_{tot}} \sum_{\mathbf{k_J}} \int_{\omega_{gap}-i\infty}^{\omega_{gap}+i\infty} d\omega \int_{cell}\int_{cell} dVdV' \left[ \text{tr}\left\{ \left[\partial_2 \hat{L}_\mathbf{k} \cdot \mathcal{G}_{\mathbf{k_J}}(\mathbf{r},\mathbf{r}',\omega)\right] \cdot \left[\partial_1 \hat{L}_\mathbf{k} \cdot \partial_\omega \mathcal{G}_{\mathbf{k_J}}(\mathbf{r}',\mathbf{r},\omega)\right]\right\}\right]$$
(D5)

For a large number of cells one can use $\frac{1}{A_{tot}} \sum_{\mathbf{k_J}} \to \frac{1}{(2\pi)^2} \iint_{BZ} d^2\mathbf{k}$, and with this substitution the above formula yields Eq. (12). This proves the equivalence between Eqs. (D1) and (12).

For local electromagnetic media we know that $\partial_i \hat{L}_\mathbf{k}$ only acts on the electromagnetic degrees of freedom: $\partial_i \hat{L} \to \partial_i \hat{N}$. Hence, the term $\text{tr}\{...\}$ in Eq. (D1) can be replaced by $\text{tr}\{\partial_2 \hat{N} \cdot \overline{\mathbf{G}}(\mathbf{r},\mathbf{r}',\omega) \cdot \partial_1 \hat{N} \cdot \partial_\omega \overline{\mathbf{G}}(\mathbf{r}',\mathbf{r},\omega)\}$ with $\overline{\mathbf{G}}$ the photonic Green function that satisfies $\hat{N}(-i\nabla) \cdot \overline{\mathbf{G}}(\mathbf{r},\mathbf{r}',\omega) = \omega \mathbf{M}(\mathbf{r},\omega) \cdot \overline{\mathbf{G}}(\mathbf{r},\mathbf{r}',\omega) + i\mathbf{1}\delta(\mathbf{r}-\mathbf{r}')$ and periodic boundary conditions over the cavity walls. Hence, we recover Eq. (29) of Ref. [26], which is a particular case of the more general Eq. (D1).

## Appendix E: Chern number for opaque-type boundaries

In this Appendix, we show that the Chern number calculated using Eq. (D1) depends critically on the boundary conditions at the cavity walls. Specifically, analogous to Ref. [31], we prove that $\mathcal{C} = 0$ when the Green function satisfies opaque-type boundary conditions on the cavity walls.

As a starting point we note that the solution of Eq. (D2) also satisfies

$$\mathcal{G}(\mathbf{r}',\mathbf{r},\omega) \cdot \left(\hat{L}(\mathbf{r},+i\overline{\nabla}) - \mathbf{M}_g(\mathbf{r})\omega\right) = i\mathbf{1}\delta(\mathbf{r}-\mathbf{r}').$$
(E1)

The arrow over the gradient operator indicates that it acts on the left, i.e., on the **r** coordinate of the Green function. To obtain this result, we introduce a $\mathcal{G}^c$ such that

$$\left(\left[\hat{L}(\mathbf{r},-i\nabla)\right]^\dagger - \mathbf{M}_g^\dagger(\mathbf{r})\omega^*\right)\cdot\mathcal{G}^c(\mathbf{r},\mathbf{r}',\omega^*) = i\mathbf{1}\delta(\mathbf{r}-\mathbf{r}'). \tag{E2}$$

Here, the dagger represents the Hermitian operator with respect to the canonical inner product, $\langle\ \rangle_{can}$. Calculating $\langle \mathcal{G}^c(\mathbf{r},\mathbf{r}'',\omega^*)\cdot\hat{\mathbf{u}}_j | (\hat{L}(\mathbf{r},-i\nabla) - \mathbf{M}_g(\mathbf{r})\omega) | \mathcal{G}(\mathbf{r},\mathbf{r}',\omega)\cdot\hat{\mathbf{u}}_i \rangle_{can}$, one may readily show that $\mathcal{G}(\mathbf{r},\mathbf{r}',\omega) = -\mathcal{G}^c(\mathbf{r}',\mathbf{r},\omega^*)^\dagger$. Transposing and conjugating Eq. (E2) and using $\mathcal{G}(\mathbf{r},\mathbf{r}',\omega) = -\mathcal{G}^c(\mathbf{r}',\mathbf{r},\omega^*)^\dagger$ one obtains Eq. (E1).

Next, we note that from Eq. (D3) the Chern number (D1) can be generally expressed as:

$$\begin{aligned}
\mathcal{C} = \lim_{A_{tot}\to\infty} \frac{-1}{A_{tot}}\mathrm{Re}\int_{\omega_{gap}-i\infty}^{\omega_{gap}+i\infty} d\omega \iint dV dV' \\
\mathrm{tr}\left\{\left[x_2\hat{L}(-i\nabla)\cdot\mathcal{G}(\mathbf{r},\mathbf{r}',\omega)\right]\cdot\left[x_1'\hat{L}(-i\nabla')\cdot\partial_\omega\mathcal{G}(\mathbf{r}',\mathbf{r},\omega)\right]\right\} \\
+\mathrm{tr}\left\{\left[\hat{L}(-i\nabla)\cdot x_2\mathcal{G}(\mathbf{r},\mathbf{r}',\omega)\right]\cdot\left[\hat{L}(-i\nabla')\cdot x_1'\partial_\omega\mathcal{G}(\mathbf{r}',\mathbf{r},\omega)\right]\right\} \\
-\mathrm{tr}\left\{\left[\hat{L}(-i\nabla)x_2\cdot\mathcal{G}(\mathbf{r},\mathbf{r}',\omega)\right]\cdot\left[x_1'\hat{L}(-i\nabla')\cdot\partial_\omega\mathcal{G}(\mathbf{r}',\mathbf{r},\omega)\right]\right\} \\
-\mathrm{tr}\left\{\left[x_2\hat{L}(-i\nabla)\cdot\mathcal{G}(\mathbf{r},\mathbf{r}',\omega)\right]\cdot\left[\hat{L}(-i\nabla')x_1'\cdot\partial_\omega\mathcal{G}(\mathbf{r}',\mathbf{r},\omega)\right]\right\}
\end{aligned} \tag{E3}$$

We would like now to integrate by parts some of the terms so that the operator $\hat{L}$ can act directly on the Green function. For example, the 4$^{th}$ term integrated by parts in **r'** gives:

$$\begin{aligned}
\iint dVdV'\, \mathrm{tr}\left\{\left[x_2\hat{L}(-i\nabla)\cdot\mathcal{G}(\mathbf{r},\mathbf{r}',\omega)\right]\cdot\left[\hat{L}(-i\nabla')x_1'\cdot\partial_\omega\mathcal{G}(\mathbf{r}',\mathbf{r},\omega)\right]\right\} \\
= \iint dVdV'\, \mathrm{tr}\left\{\left[x_2\hat{L}(-i\nabla)\cdot\mathcal{G}(\mathbf{r},\mathbf{r}',\omega)\cdot\hat{L}(+i\overleftarrow{\nabla}')\right]\cdot\left[x_1'\cdot\partial_\omega\mathcal{G}(\mathbf{r}',\mathbf{r},\omega)\right]\right\}
\end{aligned}. \tag{E4}$$

The crucial point – discussed in detail in Ref. [31] for the case of local electromagnetic media – is that the integration by parts is only possible for certain "opaque-type" boundary conditions. Specifically, if a certain state vector $\mathbf{Q}$ satisfies the relevant boundary conditions then the integration by parts requires that $x_i \mathbf{Q}$ satisfies the same boundary conditions [31]. Clearly, periodic boundary conditions are not "opaque", while, for example, a perfect electric conductor boundary is opaque.

For opaque-type boundary conditions Eq. (E3) yields:

$$\mathcal{C} = \lim_{A_{tot} \to \infty} \frac{-1}{A_{tot}} \mathrm{Re} \int_{\omega_{gap}-i\infty}^{\omega_{gap}+i\infty} d\omega \iint dVdV' x_2 x_1' \times$$

$$\left[ \mathrm{tr}\left\{ \left[\hat{L}(-i\nabla) \cdot \mathcal{G}(\mathbf{r},\mathbf{r}',\omega)\right] \cdot \partial_\omega \left[\hat{L}(-i\nabla') \cdot \mathcal{G}(\mathbf{r}',\mathbf{r},\omega)\right]\right\} \right.$$

$$+ \mathrm{tr}\left\{ \left[\mathcal{G}(\mathbf{r},\mathbf{r}',\omega) \cdot \hat{L}(+i\bar{\nabla}')\right] \cdot \partial_\omega \left[\mathcal{G}(\mathbf{r}',\mathbf{r},\omega) \cdot \hat{L}(+i\bar{\nabla})\right]\right\} \quad (E5)$$

$$- \mathrm{tr}\left\{ \left[\mathcal{G}(\mathbf{r},\mathbf{r}',\omega)\right] \cdot \partial_\omega \left[\hat{L}(-i\nabla') \cdot \mathcal{G}(\mathbf{r}',\mathbf{r},\omega) \cdot \hat{L}(+i\bar{\nabla})\right]\right\}$$

$$\left. - \mathrm{tr}\left\{ \left[\hat{L}(-i\nabla) \cdot \mathcal{G}(\mathbf{r},\mathbf{r}',\omega) \cdot \hat{L}(+i\bar{\nabla}')\right] \cdot \partial_\omega \mathcal{G}(\mathbf{r}',\mathbf{r},\omega)\right\} \right]$$

Now, we can use Eqs. (D2) and (E1) to get rid of the $\hat{L}$ operator. It can be easily checked that the $\delta$-function terms either vanish or cancel out after integration in $\omega$. Hence, we are left with:

$$\mathcal{C} = \lim_{A_{tot} \to \infty} \frac{-1}{A_{tot}} \mathrm{Re} \int_{\omega_{gap}-i\infty}^{\omega_{gap}+i\infty} d\omega \iint dVdV' x_2 x_1' \times$$

$$\left[ 2\,\mathrm{tr}\left\{ \left[\omega \mathbf{M}_g(\mathbf{r}) \mathcal{G}(\mathbf{r},\mathbf{r}',\omega)\right] \cdot \partial_\omega \left[\omega \mathbf{M}_g(\mathbf{r}') \mathcal{G}(\mathbf{r}',\mathbf{r},\omega)\right]\right\} \right.$$

$$- \mathrm{tr}\left\{ \left[\mathbf{M}_g(\mathbf{r}) \mathcal{G}(\mathbf{r},\mathbf{r}',\omega)\right] \cdot \partial_\omega \left[\omega^2 \mathbf{M}_g(\mathbf{r}') \mathcal{G}(\mathbf{r}',\mathbf{r},\omega)\right]\right\} \quad (E6)$$

$$\left. - \mathrm{tr}\left\{ \left[\omega^2 \mathbf{M}_g(\mathbf{r}) \mathcal{G}(\mathbf{r},\mathbf{r}',\omega)\right] \cdot \partial_\omega \mathbf{M}_g(\mathbf{r}') \mathcal{G}(\mathbf{r}',\mathbf{r},\omega)\right\} \right]$$

Straightforward simplifications show that the integrand vanishes. Therefore, it follows that when the Green function satisfies opaque-type boundary conditions then $\mathcal{C} = 0$.